\documentclass[journal,comsoc]{IEEEtran}

\usepackage{graphicx}
\graphicspath{{./figures/}}
\DeclareGraphicsExtensions{.eps}

\usepackage{algorithm}
\usepackage{algorithmic}
\usepackage{amsmath}
\usepackage{amsfonts}
\usepackage{amssymb}
\usepackage{array}
\usepackage{booktabs}
\usepackage{cite}
\usepackage{color}
\usepackage{enumitem}
\usepackage{multirow}
\usepackage{setspace}
\usepackage{subfigure}
\usepackage{url}

\allowdisplaybreaks[1]

\ifCLASSOPTIONcompsoc
  \usepackage[caption=false,font=normalsize,labelfont=sf,textfont=sf]{subfig}
\else
  \usepackage[caption=false,font=footnotesize]{subfig}
\fi

\newtheorem{lemma}{Lemma}

\newtheorem{theorem}{Theorem}

\newtheorem{remark}{\textbf{Remark}}

\begin{document}

\title{Online VNF Chaining and Predictive Scheduling: 
Optimality and Trade-offs}

\author{
Xi Huang,~\IEEEmembership{Student Member,~IEEE,} 
Simeng Bian,~\IEEEmembership{Student Member,~IEEE,} 
\\
Xin Gao,~\IEEEmembership{Student Member,~IEEE,}
Weijie Wu,~\IEEEmembership{Member,~IEEE,}
Ziyu Shao$^*$,~\IEEEmembership{Senior Member,~IEEE,}
\\
Yang Yang,~\IEEEmembership{Fellow,~IEEE,}
John C.S. Lui,~\IEEEmembership{Fellow,~ACM/IEEE}%
\thanks{$^*$ The corresponding author of this work is Ziyu Shao.}
\thanks{X. Huang, S. Bian, X. Gao, Z. Shao, and Y. Yang are with the School of Information Science and Technology, ShanghaiTech University, Shanghai 201210, China. (E-mail: \{huangxi, biansm, gaoxin, shaozy, yangyang\}@shanghaitech.edu.cn)}%
\thanks{W. Wu is a Research Fellow in ZeroEx Inc. (E-mail: wuwjpku@gmail.com)}%
\thanks{J. C.S. Lui is with the Department of Computer Science and Engineering, Chinese University of Hong Kong (CUHK), Hong Kong. (E-mail: cslui@cse.cuhk.edu.hk)}%
}

\maketitle

\begin{abstract}
For NFV systems, the key design space includes the function chaining for network requests and the resource scheduling for servers. 
The problem is challenging since NFV systems usually require multiple (often conflicting) design objectives and the computational efficiency of real-time decision making with limited information. 
Furthermore, 
the benefits of predictive scheduling to NFV systems still remain unexplored. 
In this paper, we propose \textit{POSCARS}, an efficient predictive and online service chaining and resource scheduling scheme that achieves tunable trade-offs among various system metrics with stability guarantee.
Through a careful choice of granularity in system modeling, we acquire a better understanding of the trade-offs in our design space. 
By a non-trivial transformation, we decouple the complex optimization problem into a series of online sub-problems to achieve the optimality with only limited information. 
By employing randomized load balancing techniques, we propose three variants of POSCARS to reduce the overheads of decision making. 
Theoretical analysis and simulations show that POSCARS and its variants require only mild-value of future information to achieve near-optimal system cost with ultra-low request response time. 
\end{abstract}


\IEEEpeerreviewmaketitle

\section{Introduction}\label{sec:introduction}
\IEEEPARstart{N}{etwork} function virtualization (NFV) is shifting the way of network service deployment and delivery by virtualizing and scaling network functions (NFs) on commodity servers in an on-demand fashion \cite{mijumbi2016network}.
As a revolutionary technique, NFV paves the way for operators towards better manageability and quality-of-service of network services.

In NFV systems, each network service is implemented as an ordered chain of virtual network functions (VNFs) that are deployed on commodity servers, {\em a.k.a.} a service chain. Along the chain, every VNF performs some particular treatment on the received requests,  
then hands over the output to the next VNF in a pipeline fashion.
To enable a network service, one needs to {\em place}, {\em activate}, and {\em chain} VNFs deployed on various servers. 
Considering the high cost of VNF migration and instantiation \cite{jiang2012joint},
VNF replacement can only be performed infrequently; that being said, when it comes to flow-scale or request-scale operations, function placement can be viewed as a static operation.
Given this fact, a natural practice is to place multiple VNFs in one server in advance, but due to hardware resource constraints (\textit{e.g.}, CPU, memory, and storage)\cite{herrera2016resource}, 
a server must carefully schedule resources among a {\em subset} of such VNFs at a particular time (\textit{i.e.}, only a subset of VNF instances can be activated on a server at a particular time). Therefore, 
with a fixed VNF placement, the activation and chaining of VNFs refer to: 1) for each server, the {\em resource allocation} to a subset of deployed VNFs subject to resource constraints; and 2) for each network service, the {\em selection} of the activated instances for its VNFs, so as to determine the sequence of instances that the requests will be treated through, \textit{a.k.a.} {\em service chaining}.

Given that VNF placement is considered static at the time scale of flow or request operations\cite{laghrissi2018survey}, 
for service chaining and resource scheduling, a natural question is: should they also be static, or dynamic? 
Static schemes have been implemented in some scenarios\cite{hantouti2018traffic}, 
but often times request traffic is highly fluctuating in both temporal and spatial dimensions\cite{kandula2009nature}.
In such cases,
static schemes may lead to workload imbalance among instances, leaving some instances overloaded and others under-utilized.
Hence, there is a huge demand to design 
an efficient and dynamic scheme that performs service chaining and resource scheduling, 
which adapts to traffic variations and achieves load balancing in real time. 
As for implementability, recent advances (\textit{e.g.}, temporal and spatial processor sharing \cite{katsikas2018metron}) have enabled real-time adjustment of resource allocation among various functions on the same server.

However, such dynamic design is non-trivial, especially in face of the complex interplay between successive VNFs and the resource contention among VNF instances on servers.
In particular, we would like to address the following challenges:

\begin{figure*}[!t]
	\centering
	\includegraphics[width=0.8\textwidth]{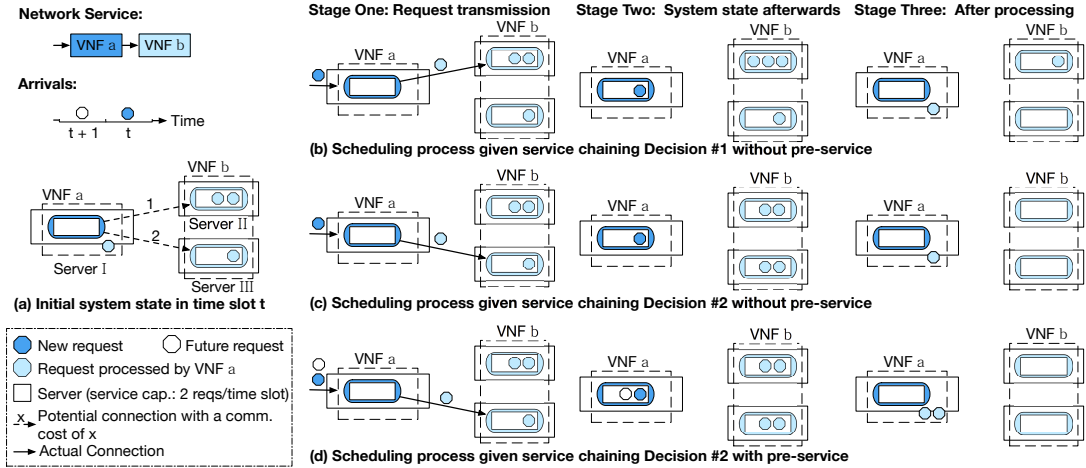}
	\caption{The system evolution in time slot $t$ under different service chaining decisions, with or without pre-service. 
	\textit{Basic settings:} There is one network service with two VNFs, \textit{i.e.}, VNF $a$ and VNF $b$. 
	VNF $a$ has one instance, while VNF $b$ has two instances.
	Every instance maintains one queue to buffer untreated requests. 
	All the instances are readily deployed, with VNF $a$'s instance on server I, while the instances of VNF $b$ on server II and III, respectively. 
	VNF $a$'s instance is potentially connected to both instances of VNF $b$.
	\textit{Initial state:} 
	one new request has arrived at time $t$ and another one will arrive at time $(t+1)$. 
	Besides, VNF $a$'s instance has one request that has been processed in time $(t-1)$ and to be sent to one of VNF $b$'s instances in time $t$. 
	\vspace*{-0.1in}
	}
	\label{motiv_example}
\end{figure*}
\setlength{\textfloatsep}{0pt}

\begin{enumerate}
	\item {\textbf {Characterization of the tunable trade-offs among various performance metrics:} }
NFV systems often have multiple optimization objectives, \textit{e.g.}, maximizing resource utilization, minimizing energy consumption, and reducing request response time. 
Different stakeholders may have different preferences over these objectives which often times conflict with each other\cite{schneider2018trade}. It is important to characterize their trade-offs to acquire a comprehensive understanding of our design space and tune the system towards the particular state that we desire. 
  	\item {\textbf {Efficient online decision making:}}
VNF request processing often requires low latency and high throughput.  
Hence, an effective dynamic scheme must also be computationally efficient, and can be adaptive to request changes. 
This is challenging not only because of the nature of the high complexity, but also that service requests arrive in an online manner, while the underlying traffic statistics are often unknown a priori. 
All these uncertainties make it  more challenging to optimize system objectives through a series of online decisions, 
not to mention that a distributed design is often preferred.
	\item \textbf{Understanding benefits of predictive scheduling:} A natural optimization of online decision making is to consider leveraging recently developed machine learning techniques \cite{box2015time}\cite{taylor2018forecasting} to predict future traffic information to reduce response time and improve quality-of-service. 
There is no free lunch, though. 
For example, in NFV-based multimedia delivery systems, multimedia service providers can predict potential requests based on the popularity of streaming contents and subscribers' preferences \cite{NetflixPred}. Based on such predictions, service providers can carry out pre-rendering or compression to optimize the quality of their services with faster responses\cite{bouten2015towards}.
Despite the wide adoption of such prediction-based approaches \cite{zhang2015proactive}\cite{nanda2016predicting}\cite{huang2019sdn}\cite{huang2019streaming}\cite{gao2019fog}, 
it still remains open what are the fundamental benefits of predictive scheduling to NFV systems, even in the presence of prediction errors. 
Answers to the questions are the key to understand whether the endeavor worthy to put on predictive VNF scheduling, 
and whether one can tolerate the worst possible case that may occur. 
\end{enumerate}

{ 
\vspace{0em}

Despite recent headway on VNF scheduling\cite{herrera2016resource}\cite{hantouti2018traffic}, as far as we are aware, 
there is still no fundamental understanding on the above questions, nor is there any strategy that can achieve the design objectives simultaneously in a fully online fashion. 
One important reason is in the difficulty of problem formulation and modeling, especially in choosing the granularity. 
If one models the system state and strategy in flow-level abstraction\cite{mohammadkhan2015virtual}, it may fall short in accurate characterization of interplay between successive VNF instances and system dynamics over time; however, if one applies fine-grained control to each request\cite{zhang2015proactive}, then the decision making will inevitably incur a rather high computational overhead. 
Such issue not only prohibits a deep understanding on system dynamics, but also prevents us from obtaining efficient and accurate strategy design.

In this paper, we overcome such difficulties by applying a number of novel techniques. Our contributions include:
\vspace{0em}

{\bf Modeling and formulation:} 
We propose a novel model that separates the granularity of system state characterization and strategy making. In particular, we develop a queuing model at the request granularity to characterize system dynamics. 
Unlike flow-level abstraction, our model require no prior knowledge on underlying flows, 
but accurately captures the interplay 
between successive instances, 
\textit{i.e.}, real-time dynamics of how requests are received, processed, and forwarded. 
As for strategy making, it is conducted at the granularity of request batch in a per-time-slot fashion to avoid the high overheads of per-request optimization.
Such a careful choice makes it possible to characterize the system dynamics and performance in a clear yet accurate way. 

{\bf Algorithm design:}  
To enable online and efficient decision making, we transform the long-term stochastic optimization problem into a series of sub-problems over time slots. 
By exploiting their unique structure, 
we propose \textit{POSCARS}, a Predictive Online Service Chaining And Resource Scheduling scheme. 
{Particularly, POSCARS includes two coupled parts. One is for the predictive scheduling of requests, while the other is for service chaining and resource allocation. The former part takes advantage of predicted information to effectively reduce request delays. Meanwhile, the latter part can incur a near-optimal system cost while stabilizing all queues in the system. Furthermore, it can also achieve a tunable control between system cost optimization and queue stability.}

{\bf Predictive scheduling:} 
To the best of our knowledge, this paper is the first to address the dynamic service chaining and scheduling problem in NFV system by jointly considering resource utilization, energy efficiency, and request latency. 
This paper is also the first to study the fundamental benefits of predictive scheduling with future information in NFV system, which extends a new dimension for NFV system design. 

{\bf Experiment verification and investigation:}  
We conduct trace-driven simulations and results show the effectiveness of POSCARS and its variants under various settings against baseline schemes, as well as the benefits of predictive scheduling in achieving ultra-low request response time. 

The rest of this paper is organized as follows. 
In Section \ref{sec: motivating example}, we show a motivating example of predictive scheduling in NFV systems. 
Section \ref{sec: problem formulation} presents our model and formulation, followed by the design and performance analysis of POSCARS and its variants in Section \ref{sec: algorithm design}. 
We show simulation results and analysis in Section \ref{sec: simulation}, then review related work in Section \ref{sec: related work}. 
Finally, Section \ref{sec: conclusion} concludes the paper.

\section{Motivating Example}  \label{sec: motivating example}
In this section, we first show a motivating example that exhibits the potential trade-off in the multi-objective optimization for different system metrics, including reduction in energy cost and communication cost, as well as shortening response times, which is mainly due to queueing delay. 
Besides, the example also explores the value of future information and the potential benefit of predictive scheduling.

We consider a time slotted NFV system, where predictive scheduling is viable, \textit{i.e.}, the request in time $(t+1)$ can be perfectly predicted, pre-generated, and pre-served by the system.\footnote{
		An example of predictive scheduling in practical systems is that Netflix predicts its users' behaviors and preloads video onto their devices\cite{NetflixPred}.
	} 
Figure \ref{motiv_example}(a) shows the basic settings and initial system state in time slot $t$. 
All VNF instances are readily deployed on servers with a fixed placement.
Each instance maintains a queue to buffer any untreated request.
Every server has a service capacity of two requests per time slot; processing a request incurs an energy cost of $1$.
Note that 1) any requests processed by VNF $a$'s instance are not counted in the queues, but readily to be sent to VNF $b$'s instances in the next time slot; 2) requests that have been processed by VNF $b$'s instances are considered finished.

In this case, there are two possible service chaining decisions,
\textit{i.e.},
forwarding the processed request from the instance of VNF $a$ to either VNF $b$'s instance on server II (Decision \#1) or server III (Decision \#2).
It takes a communication cost of $1$ to forward the request to VNF $b$'s instance on server II. 
The communication cost is $2$ to the other instance of VNF $b$ on server III.
 
Our goal is to choose a service chaining decision in time $t$ that jointly minimizes the total energy cost, total communication cost, and the total residual backlog size at the end of time $t$. 
\footnote{By applying \textit{Little's law}, a short queue length implies short queueing delay or short response time.}   
Figures \ref{motiv_example}(b) - \ref{motiv_example}(d) compare the scheduling processes under different service chaining decisions. 

In Figure \ref{motiv_example}(b), the new request in time $t$ is admitted, while the processed request is forwarded to the instance of VNF $b$ on server II. 
Although incurring a low communication cost of $1$, such a decision also leads to imbalanced queue loads among VNF $b$'s instances.
Note that every server can serve at most two requests per time slot.
Hence, servers will then process four requests in total, including the new request on server I, two requests on server II, and another one on server III. 
The processing incurs a total energy cost of $4$. 
After processing, VNF $b$'s instance on server II still has one untreated request in its backlog. 
Thus Decision \#1 \textit{incurs a total cost of $5$ on energy and communication, with a residual backlog size of $1$}.

On the other hand, when the processed request is forwarded to the instance of VNF $b$ on server III,
the decision incurs a high communication cost of $2$ but results in balanced queue loads among VNF $b$'s instances. 
Servers will process five requests in total, including the new request on server I, and the rest from server II and III. 
The processing incurs a total energy cost of $5$.
After processing, there are no untreated request left in the backlogs. 
Decision \#2 incurs \textit{a higher total cost of $7$ on energy and communication, but with no residual backlogs}.

\textbf{Insight 1:}
Figures \ref{motiv_example}(b) and \ref{motiv_example}(c) show that we cannot achieve the optimal values for different system metrics simultaneously, 
\textit{i.e.}, there is a potential trade-off between optimizing the total system cost and reducing the total queue length. 

Additionally, we find that server I is under-utilized in both Figures \ref{motiv_example}(b) and \ref{motiv_example}(c),
because VNF $a$'s instance only receives and handles the new request at time $t$. 
In fact, Figure \ref{motiv_example}(d) shows that we can exploit the spare processing power on server I by pre-admitting and pre-serving the future request. 
Consequently, 
we can shorten the response time for the future request by incurring one more energy cost in time $t$. 
Note that pre-service does not introduce extra energy cost but actually pays it beforehand. 
The reason is that even without pre-service, we still have to pay one energy cost in the subsequent time slots after the future request arrives.  

\textbf{Insight 2:}
By utilizing servers' spare processing power and paying system cost in advance, 
predictive scheduling can effectively shorten response times of future requests. 

To characterize the non-trivial trade-off and exploit the power of predictive scheduling in NFV systems, we present our formulation in the next section.

\begin{figure}[!t]
\centering
\includegraphics[scale=.24]{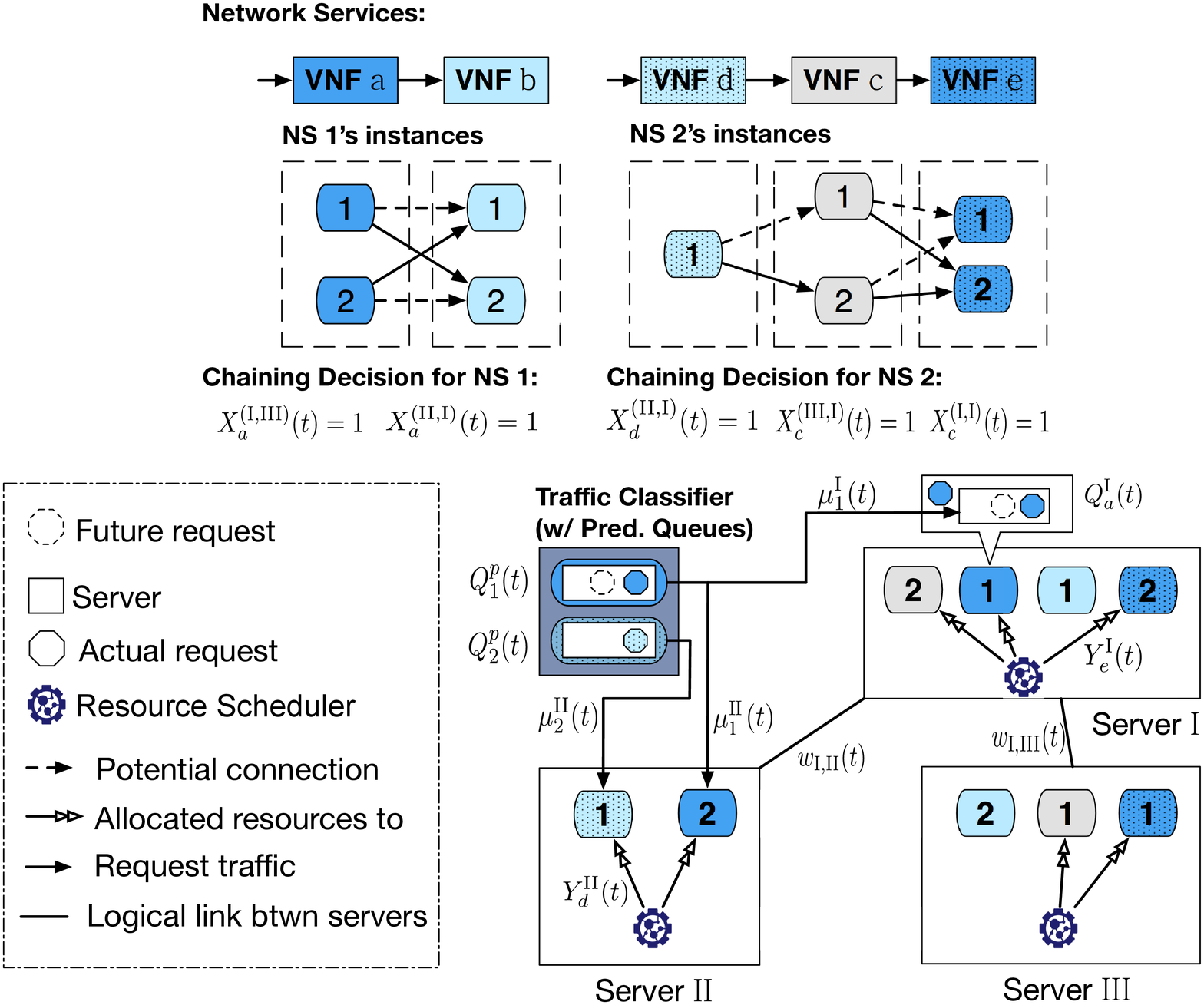}
\caption{An instance of our system model. There are two network services (NS 1 and NS 2) with their VNF instances deployed on servers I, II, and III. At the beginning of time slot $t$, the traffic classifier admits and push requests to the queues $Q^p_1(t)$ and $Q^p_2(t)$ with respect to their requested services. For each instance, based on its server's resource scheduling, it serves requests from its processing queue and forwards requests to its next VNF instances, \textit{e.g.}, instance of VNF $a$ on server I to instance of VNF $b$ on server III.}
\label{system_model}
\end{figure}
\setlength{\textfloatsep}{0pt}

\section{Problem Formulation}  \label{sec: problem formulation}
We consider a time slotted NFV system, 
where virtualized network functions (VNF) are instantiated, deployed over a substrate network, and chained together to deliver numbers of network services. 
Upon the arrival of new network service requests, each VNF processes and hands over requests to its following VNF in a pipeline fashion. 
All requests are assumed homogeneous; \textit{i.e.}, each request is assumed to have equal size and require the same amounts of computation to be processed.
We show an instance of our system model in Figure \ref{system_model} and summarize main notations in Table \ref{notations}.
More details of service chaining can be found in IETF RFC-7665 \cite{halpern2015service}.

\begin{table}
	\caption{Main Notations} \label{notations}
	\begin{tabular}{l|l}
	\toprule
	\multicolumn{2}{c}{\textbf{Substrate Network Model}} \\
	\hline
	$\mathcal{S}$ & The set of servers that host VNFs \\
	\hline
	\multirow{2}*{$w_{s',s}(t)$} & Communication cost of sending one request\\
	& from server $s'$ to server $s$ \\
	\hline
	\multirow{2}*{$c_{s,i}$} & The capacity of type-$i$ resource type on \\
	& server $s$ \\ 
	\hline
	$\lambda_{s,i}$ & The unit cost of type-$i$ resource on server $s$ \\
	\hline
	\hline
	\multicolumn{2}{c}{\textbf{Network Service Model}} \\
	\hline
	$K$ & Number of network services \\
	\hline
	$\mathcal{L}_k$ & Chain length of network service $k$ \\
	\hline
	$\mathcal{F}$ & The set of virtual network functions (VNFs)\\
	\hline
	$\mathcal{F}_{in}$ & The set of all ingress VNFs \\
	\hline
	$\mathcal{F}_{nt}$ & The set of all non-terminal VNFs \\
	\hline
	$f_{k,j}$ & The $j$-th VNF in network service chain $k$ \\
	\hline
	$k_{f}$ & The network service that contains VNF $f$ \\
	\hline
	$p(f)$ & Previous VNF of $f$ in service chain $k_f$ \\
	\hline
	$n(f)$ & Next VNF of $f$ in service chain $k_f$ \\
	\hline
	\multirow{2}*{$\theta_{f,i}$} & Number of requests processed by unit of \\
	& type-$i$ resource \\
	\hline
	\hline
	\multicolumn{2}{c}{\textbf{Deployment Model}} \\
	\hline
	$\mathcal{S}_f$ & The set of all servers that host $f$'s instances \\
	\hline
	\multirow{2}*{$\mathcal{F}_s$} & The set of VNFs with instances residing on \\
	& server $s$ \\
	\hline
	\hline
	\multicolumn{2}{c}{\textbf{System Dynamics}} \\
	\hline
	\multirow{2}*{$A_k(t)$} & Number of new requests for network \\
	& service $k$ arriving in time $t$ \\
	\hline
	\multirow{2}*{$Q^{(d)}_k(t)$} & Number of untreated requests for network \\
	& service $k$ in the next $d$-th slot from time $t$ \\
	\hline
	\multirow{2}*{$Q^{p}_k(t)$} & The prediction queue length for \\
	& network service $k$ in time $t$ \\
	\hline
	\multirow{2}*{$Q^{s}_f(t)$} & Queue length of VNF $f$'s instance on \\
	& server $s$ in time $t$ \\
	\hline
	\multirow{3}*{$B^{s}_{f}(t)$} &
	Number of processed requests by VNF $f$'s \\
	& instance on server $s$ in time $(t-1)$, and to \\
	& be sent to next VNF in time $t$ \\
	\hline
	\multirow{2}*{$\delta_{k}(t)$} & Total number of admitted requests for network \\
	& service $k$ \\
	\hline
	$\delta^{(d)}_{k}(t)$ & Number of admitted requests from $Q^{(d)}_{k}(t)$ \\
	\hline
	\hline
	\multicolumn{2}{c}{\textbf{Scheduling Decisions}} \\
	\hline
	\multirow{2}*{$\mu^{s}_{k}(t)$} & Number of admitted requests for service $k$ \\
	& onto server $s$ \\
	\hline
	\multirow{4}*{$X^{(s', s)}_{f}(t)$} & $X^{(s', s)}_{f}(t) = 1$ if the instance of $n(f)$ on \\
	& server $s$ is selected to receive the processed \\
	& requests from $f$'s instance on server $s'$ and \\
	& $0$ otherwise \\
	\hline
	\multirow{2}*{$Y^{s}_{f}(t)$} & the vector of allocated resources on server $s$  \\
	& VNF $f$'s instance \\
	\hline
	\hline
	\multicolumn{2}{c}{\textbf{System Objectives}} \\
	\hline
	$m(t)$ & Total communication cost in time $t$ \\
	\hline
	$g(t)$ & Total computation cost in time $t$ \\
	\hline
	$h(t)$ & Weighted total queue length in time $t$ \\
	\bottomrule
	\end{tabular}
\end{table}

\subsection{Substrate Network Model}

We consider the substrate network with a set $\mathcal{S}$ of heterogeneous servers.
On each server $s$, we consider $R$ types of resources, \textit{e.g.}, GPU \cite{zhang2018g}, CPU cores \cite{katsikas2018metron}, and cache \cite{tootoonchian2018resq}. 
The $i$-th resource type has a capacity of $c_{s, i}$ and a unit cost of $\lambda_{s, i}$.  
We denote the resource capacity vector $[c_{s,i}]_{i=1}^{R}$ by $\boldsymbol{c}_s$, and the resource unit cost vector $[\lambda_{s, i}]_{i=1}^{R}$ by $\boldsymbol{\lambda}_s$.

For every server pair $(s', s)$, 
we use $w_{s', s}(t)$ to denote the communication cost of transferring a request between the servers in time $t$, \textit{e.g.}, the number of hops or round-trip times.  
If two servers are not reachable from each other in time $t$, then we set $w_{s', s}(t) = +\infty$. 
The set of all communication cost $[w_{s', s}(t)]_{s', s}$ in time slot $t$ is denoted by $\boldsymbol{w}(t)$.

\subsection{Network Service Model} 
There are $K$ network services and a set $\mathcal{F}$ of VNFs. Each network service $k$ is represented by a chain of $L_{k}$ ordered VNFs, 
wherein the $j$-th VNF is denoted by $f_{k, j}$. 
To avoid triviality, we assume that $L_{k} \ge 2$ for every network service $k$. Note that $L_{k}$ is a constant and usually not very large\cite{han2015network}. 
We regard the same VNF that appears in different service chains as distinct VNFs.
In practice, one can set up multiple queues on one VNF instance to buffer requests for different services and 
map each queue to one VNF instance in our model.

Next, we use $\mathcal{F}_{in}$ to denote $\{f_{k, 1}\}_{k=1}^{K}$, \textit{i.e.}, the set of ingress VNFs of all network services,
and $\mathcal{F}_{nt}$ to denote the set of non-terminal VNFs of all network services.
For every VNF $f \in \mathcal{F}$, we denote its network service by $k_f$. 
If $f \notin \mathcal{F}_{in}$, \textit{i.e.}, not the first VNF of its network service,
then we denote its previous VNF by $p(f)$; 
likewise, if $f \in \mathcal{F}_{nt}$, \textit{i.e.}, not a terminal VNF, then we denote its next VNF by $n(f)$. 
\subsection{Deployment Model}
In practice, due to request workload changes, it's common to provide multiple instances for every VNF, 
	encapsulate the instances into containers, 
	and distribute them on servers for better load balancing and fault tolerance \cite{woo2018elastic}. 
We assume that each VNF has at most one instance on each server but it can have multiple instances on different servers. 
The placement of VNF instances is assumed to be pre-determined by adopting VNF placement schemes similar to existing ones\cite{zhang2018placement, zhang2017joint, cohen2015near, sang2017provably}. 
Depending on the placement, the instances required by each service are not necessarily readily available on the same server.
Note that our model can be further extended to cases with each VNF having multiple instances on the same server.  

For VNF $f \in \mathcal{F}$, we use $\mathcal{S}_{f}$ to denote the set of servers that host $f$'s instances. 
Correspondingly, each server $s$ hosts a subset $\mathcal{F}_s \subseteq \mathcal{F}$ of VNFs. 
Every instance maintains one queue to buffer its relevant requests. 
For example, if VNF $f$ has one instance on server $s$, then the instance has a queue of size $Q^{s}_{f}(t)$ at the beginning of time slot $t$. 
Instead of individual queues, one can also implement a shared public queue among instances of the same VNF. 
All requests from preceding VNF's instances are firstly forwarded and buffered in the public queue. 
These buffered requests are then rescheduled to one or more idle or least loaded instances.
Such a way brings more flexibility so that requests can avoid the potential long queueing delay on individual instances. 
However, it requires additional physical storage and communication cost due to additional rescheduling. 
The choice depends on the trade-off made by system designers. Here we adopt the queueing model for each individual instance.

\subsection{Predictive Request Arrival Model}

For network service $k$, we use $A_{k}(t)$ ($\le a_{max}$ for some constant $a_{max}$) to denote the number of its new requests that arrive in time slot $t$, and independent over time slots. 
In practice, considering the statefulness of VNFs, requests may be aggregated and scheduled in the unit of flow.
Our model captures the system dynamics at a finer granularity than the flow-level abstraction, and can be further extended to the case with correlation between requests. 

Next, we consider a system which can predict and pre-serve future request arrivals for network services in a finite number of time slots ahead. 
Though the technique and analysis of prediction is still under active development\cite{zhang2015proactive, petrov2018mathematical, nanda2016predicting}, 
we do not assume any particular prediction technique in this paper. 
Instead, we assume the prediction as the output from other standalone predictive modules, and investigate the \textit{fundamental} benefits by acquiring and leveraging such future information and the risks induced by mis-prediction.
Note that such an assumption is valid to approximate practical scenarios where short-term prediction is viable. 
For example, Netflix promotes its quality-of-experience (QoE) by predicting user demand and network conditions, then prefetching video frames onto user devices\cite{NetflixPred}.

We assume that for network service $k$, the system has perfect access to its future requests  
in a prediction window of size $D_k$ ($\le D$ for some constant $D$), denoted by $\{A_k(t+1), \dots, A_k(t+D_k)\}$.
In practice, however, such prediction may be error-prone; we shall evaluate the impact of mis-prediction in the simulation.
With pre-service, some future requests may have been admitted into or even pre-served before time $t$, thus we use $Q^{(d)}_k(t)$ ($0\leq d\leq D_k$) to denote the number of untreated requests in slot $(t+d)$ at time $t$, such that 
\begin{equation}
	\begin{array}{c}
		0 \leq Q^{(d)}_k(t) \leq A_k(t+d).
	\end{array}
\end{equation}
Note that $Q^{(0)}_k(t)$ denotes the number of untreated requests that arrive at time $t$. 
Therefore, the total number of untreated requests for service $k$ is 
$Q^p_k(t) = \sum_{d=0}^{D_k} Q^{(d)}_k(t)$. 
Here we can treat $Q^{p}_k(t)$ as a virtual prediction queue that buffers untreated future requests for network service $k$. 
In practice, the prediction queues can be hosted on servers or storage systems in proximity to the request traffic classifier \cite{sheoran2017contain}. 
To simplify notations, we use $\boldsymbol{Q}(t)$ to denote the vector of all queues' length $\{Q^p_k(t)\}_{k=1}^{K}$ and $\{Q^{s}_{f}(t)\}_{s \in \mathcal{S}, f \in \mathcal{F}_{s}}$.

\subsection{System Workflow and Scheduling Decisions}

\textbf{System Workflow:}
At the beginning of each time slot $t$, system components (including traffic classifier, VNF instances, and servers) collect relevant system dynamics to decide request admission, service chaining, and resource allocation. 
According to the decisions, 
the traffic classifier admits new requests for different network services. 
VNF instances steer the requests which are processed in time slot $(t-1)$ to their next VNF's instances. 
Meanwhile, every server allocates the resources to its resident VNF instances\cite{katsikas2018metron}. 
The instances then process the requests from their respective queues. 
At the end of time slot, the prediction window moves one slot ahead.

In the above process, we need to consider three kinds of scheduling decisions.

\textbf{i) Admission Decision:}
For every network service, 
the traffic classifier decides the number of  
untreated newly arriving and future requests, to admit into the system.
Particularly, for a network service $k$ and its respective ingress VNF $f$, 
the classifier decides
 $\mu^{s}_{k}(t)$, \textit{i.e.},
 the number of admitted requests to $f$'s instance on server $s \in \mathcal{S}_{f}$. 
We use $\delta_{k}(t)$ to denote the total number of admitted requests from prediction queue $Q^{p}_{k}(t)$. 
These admitted requests should include at least all the untreated requests that actually arrive, while not exceeding $Q^{p}_{k}(t)$, \textit{i.e.}, in time slot $t$ and for $k = 1, \dots, K$,

\begin{equation}\label{constraint_mu}
	\setstretch{0.5}
	\begin{array}{c}
		\displaystyle
		Q^{(0)}_{k}(t) \le \delta_{k}(t) \triangleq \sum_{s \in \mathcal{S}_{f_{k,1}}} \mu^{s}_{k}(t) \le Q^{p}_{k}(t).
	\end{array}
\end{equation}
Note that requests are admitted in a fully-efficient manner \cite{huang2016predictive}. In other words, by admitting $\delta^{(d)}_{k}(t)$ untreated requests from $Q^{(d)}_{k}(t)$ for $0 \le d \le D_{k}$, 
the allocation should ensure a total number of $\delta_{k}(t)$ requests to be admitted, \textit{i.e.},
\begin{equation}
		\sum_{d=0}^{D_{k}} \delta^{(d)}_{k}(t) = \delta_{k}(t)
		\enskip
		\forall k \in \{1, \dots, K\}.
\end{equation}
The untreated request backlog $Q^{(d)}_{k}(t)$ evolves as follows,
\begin{equation}
	\begin{array}{c}
		\displaystyle
		Q^{(d)}_{k}(t+1) = \left[ Q^{(d+1)}_{k}(t) - \delta^{(d+1)}_{k}(t)  \right]^{+}, \!\!
		\enskip
		\forall~ d \in [0, D_{k} - 1].
	\end{array}
\end{equation}
while $Q^{(D_{k})}_{k}(t+1) = A_{k}(t+D_{k}+1)$, where we define $[x]^{+} \triangleq \text{max}\{x, 0\}$.
We denote all admission decisions by $\boldsymbol{\mu}(t)$. 

\textbf{ii) Service Chaining Decision:}
Given a non-terminal VNF $f$, we denote $X^{(s', s)}_{f}(t) \in \{0, 1\}$ as the service chaining decision at time $t$.
We consider the case when VNF $f$ and its next VNF $n(f)$ have instances on server $s'$ and $s$, respectively.
The decision with value $1$ indicates the processed requests from VNF $f$'s instance on server $s'$ will be sent to $n(f)$'s instance on server $s$,
and $0$ otherwise.
To ensure that every instance has a target instance to send its requests, we have
\begin{equation}\label{constraint_x}
	\begin{array}{c}
		\displaystyle
		\sum_{s \in \mathcal{S}_{n(f)}} 
		X^{(s', s)}_{f}(t) = 1,
		\enskip\enskip
		\forall s' \in \mathcal{S}_{f}, \ \forall t.
	\end{array}
\end{equation}
On the other hand, if VNF $f$ (or its next VNF) has no instances on server $s$ (or $s'$), then $X^{(s, s')}_{f}(t) = 0$ in each time slot $t$.
Note that dynamic request steering can be implemented by adopting VNFs-enabled SDN switches \cite{hsieh2017nf}. 
We denote all chaining decisions by $\boldsymbol{X}(t)$.

\textbf{iii) Resource Scheduling Decision:}
For each server $s$ and VNF $f \in \mathcal{F}_{s}$, we define $Y^{s}_{f}(t) \in \mathbb{Z}^{R}_{+}$ as the allocated resource vector to $f$'s instance. 
To ensure any allocation with at least one CPU core and other resources, or without any resources at all, 
we restrict the choice of $Y^{s}_{f}(t)$ to a finite set of options $\mathcal{O}_{f}$.
Note that $\emptyset \in \mathcal{O}_{f}$ for all $f$, \textit{i.e.}, the option of no resource allocation is always available. 
Besides, the total allocated resources should not exceed server $s$'s resource capacity, \textit{i.e.},
\begin{equation}\label{constraint_y}
	\begin{array}{c}
		\displaystyle
		\sum_{f \in \mathcal{F}_{s}} Y^{s}_{f}(t) \preceq \boldsymbol{c}_s,
		\enskip \forall s \in \mathcal{S}, ~ \forall t.
	\end{array}
\end{equation}
Note that $Y^{s}_{f}(t) = \emptyset$ for all the time if $f \notin \mathcal{F}_{s}$.
Given resource allocation $Y^{s}_{f}(t)$, 
the instance can process and forward at most $\phi_f(Y^{s}_{f}(t))$ requests, 
where $\phi_{f}(\cdot)$ is assumed to be estimated from system logs.
Due to time slot length limit, a VNF instance can't process too many requests and thus we assume $\phi_{f}(\cdot) \le \phi_{max}$ for some constant $\phi_{max}$.
We denote all allocation decisions by $\boldsymbol{Y}(t)$.

\subsection{System Workflow and Queueing Dynamics}
\begin{figure}[!t]
    \begin{center}
        \includegraphics[scale=.28]{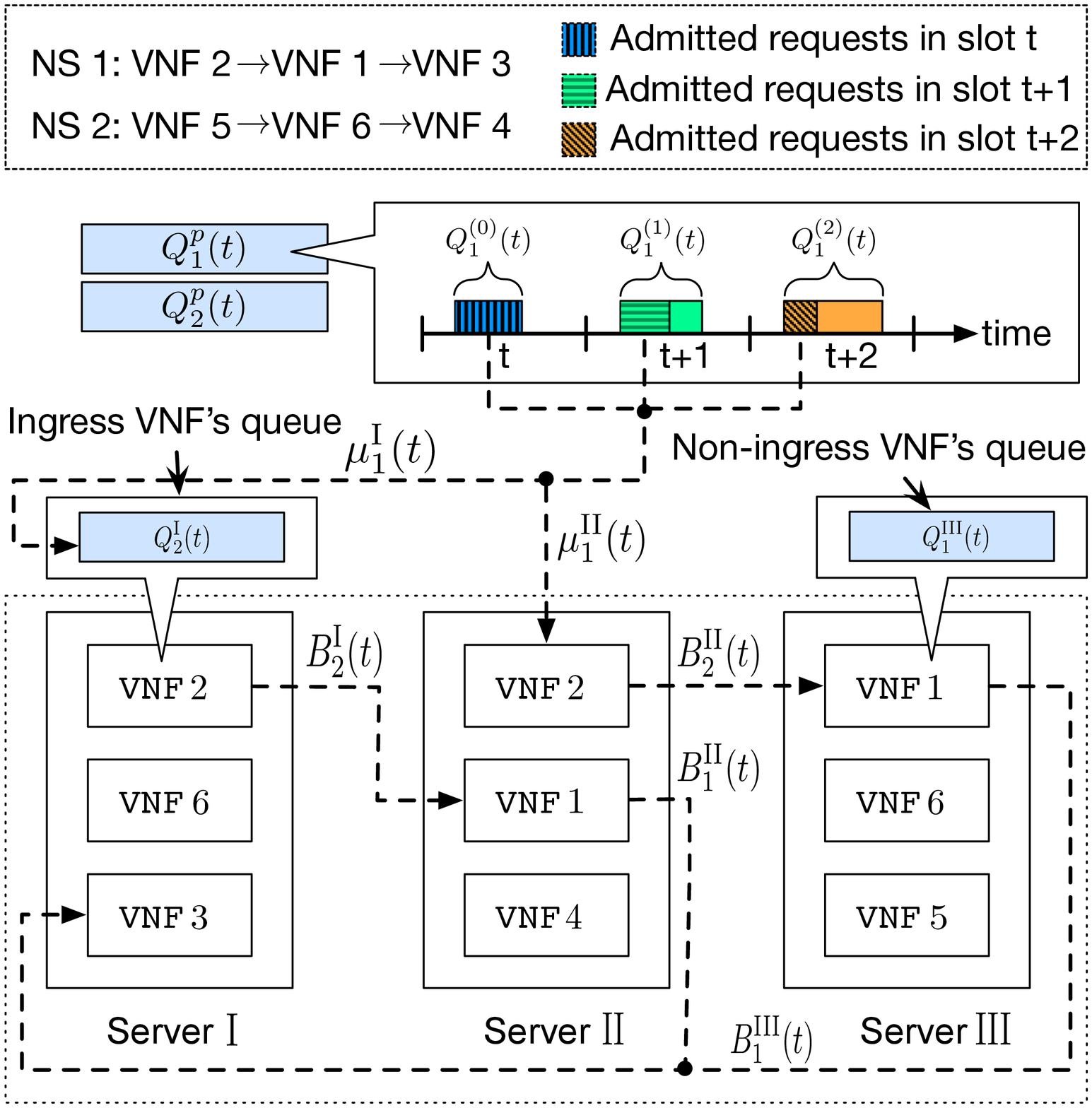}
    \end{center}
    \caption{ 
    	An instance of queueing model with a lookahead window size of two. 
    }
    \label{queue_model}
\end{figure}
\setlength{\textfloatsep}{0pt}
In time slot $t$, the system workflow proceeds as follows. 
At the beginning of time slot $t$, system components (including traffic classifier, VNF instances, and servers) collect all available system dynamics to make request admission, service chaining, and resource allocation decisions $[\boldsymbol{\mu}(t), \boldsymbol{X}(t), \boldsymbol{Y}(t)]$. 
According to the decisions, 
traffic classifier admits new requests for different network services. 
VNF instances steer the requests which are processed in time slot $(t-1)$ to their next VNF's instances. 
Meanwhile, every server allocates the resources to its resident VNF instances. 
The instances then process the requests from their respective queues. 
At the end of time slot $t$, the prediction window for each network service $k$ moves one slot ahead. 
Thus given $\mu_{k}(t)$, prediction queue $Q^{p}_{k}(t)$ is updated as follows
\begin{equation}\label{q_update_pred}
		Q^{p}_{k}(t+1) = [Q^{p}_{k}(t) - 
		\sum_{s \in \mathcal{S}_{f}} \mu^{s}_{k}(t)
		]^{+} + A_{k}(t + D + 1).
\end{equation}
With the above workflow, we have the subsequent queueing dynamics for different VNF instances. 

\textbf{Instances of Ingress VNFs:} 
For every network service $k$ and its respective ingress VNF $f$, 
there are $\mu^{s}_{k}(t)$ admitted requests to $f$'s instance on server $s \in \mathcal{S}_{f}$.
Accordingly, the update function for queue length $Q^{s}_{f}(t)$ is 
\begin{equation}
	\begin{array}{c}\label{q_update_ingr}
		\displaystyle
		Q^{s}_{f}(t+1) = 
		\left[
		Q^{s}_{f}(t) - \phi_f(Y^{s}_{f}(t))\!\! + 
		\mu^{s}_{k_{f}}(t) 
		\right]^{+}.
	\end{array}
\end{equation}

\textbf{Instances of Non-Ingress VNFs:} 
For the instance of VNF $f \notin \mathcal{F}_{in}$ on server $s$,
if $X^{(s',s)}_{p(f)}(t) = 1$, then the instance
will receive processed requests from the instance of VNF $p(f)$ on server $s'$;
otherwise, the instance will receive no new requests. 
then the queueing update function is given by
\begin{equation}\label{q_update_non_ingr}
		Q^{s}_{f}(t+1) \le 
		\Big[Q^{s}_{f}(t) - \phi_{f}(Y^{s}_{f}(t))\!\! + \!\!\!\!\!\!\!\!
		\sum_{s' \in \mathcal{S}_{p(f)}} \!\!\!
		X^{(s', s)}_{p(f)}(t) \!\cdot\! B^{s'}_{p(f)}(t)
		\Big]^{+},
\end{equation}
where
$
	B^{s'}_{p(f)}(t) \triangleq 
	\phi_{p(f)}\big( Y^{s'}_{p(f)}(t-1) \big),
$
\textit{i.e.}, the allocated service rate for the instance of $p(f)$ on server $s'$ in time $(t-1)$. 
The inequality is due to that the actual number of untreated requests may be less than the service rate in time $(t-1)$.
All requests processed by the last instances of service chains are considered finished. 
The vector $[B^{s}_{f}(t)]_{s, f}$ is denoted by $\boldsymbol{B}(t)$.
Figure \ref{queue_model} shows an example of our queue model, in which there are two network services that require six types of VNF whose instances are hosted on three servers; each of the network services has a prediction window of size two. In Figure \ref{queue_model}, we show how requests are admitted and transferred between successive queues for the first network service (NS $1$) in time $t$, given admission decision $\mu^{\text{I}}_{1}(t)$, $\mu^{\text{II}}_{1}(t)$, and chaining decision 
    	$X^{(\text{I}, \text{II})}_{2}= X^{(\text{II}, \text{III})}_{2}=
    	X^{(\text{II}, \text{I})}_{1}=
    	X^{(\text{III}, \text{I})}_{1}=
    	1$.
\vspace{-0.3em}
\subsection{Optimization Objectives}
\textbf{Communication Cost:}
Recall that transferring a request over link $(s', s)$ incurs a communication cost $w_{s', s}(t)$, \textit{e.g.}, the number of hops or round-trip times.
Low communication cost are highly desirable for  responsiveness of requests. 
In time slot $t$, 
    given the service chaining decisions, 
    the communication cost between server $s$ and $s'$ is 
    \begin{equation}\label{def_comm_cost_server_pair}
    	\setstretch{1.2}
    		m_{s', s}(t) 
    		\triangleq 
    		\hat{m}_{s', s}(\boldsymbol{X}(t))
    		= \!\!\!
    		\sum_{f \in \mathcal{F}_{nt}} 
    		B^{s'}_{f}(t) X^{(s', s)}_{f}(t) w_{s', s}(t).
    \end{equation}    
    where $w_{s', s}(t)$ denotes the communication cost of transferring a request between servers $s'$ and $s$ in time $t$.
    Then the total communication cost in time $t$ is given by
    \begin{equation}\label{def_comm_cost_total}
        \setstretch{1.2}
		m(t) \triangleq \hat{m}\left( \boldsymbol{X}(t) \right) 
		= 
		\sum_{s', s \in \mathcal{S}} 
		\hat{m}_{s', s}(\boldsymbol{X}(t)).	
    \end{equation}   
    
\textbf{Energy Cost:}
Efficient resource utilization for servers is another important objective to achieve in NFV systems \cite{xu2016demystifying}. 
Given the resource allocation $Y^{s}_{f}(t)$, we define the corresponding energy cost in time $t$ as 
	$\boldsymbol{\lambda}^{T} Y^{s}_{f}(t)$, 
    where $\boldsymbol{\lambda} \in \mathbb{Z}^{R}_{+}$ is a constant vector, with each entry $\lambda_{i}$ as the unit cost of $i$-th type of server resources. 
The total energy cost in time $t$ is
    \begin{equation}\label{def_comp_cost_total}
    \begin{split}
		\displaystyle g(t) \triangleq \hat{g}(\boldsymbol{Y}(t)) = 
		\sum_{s \in \mathcal{S}} \sum_{f \in \mathcal{F}_{s}}
		\boldsymbol{\lambda}^{T} Y^{s}_{f}(t).
	\end{split}
	\end{equation}
   
\textbf{Queue Stability:}
    Considering the responsiveness of requests and scarcity of computational resources such as memory and cache,
    it is also imperative to ensure that no queues would be overloaded.
    We denote the weighted total queue length in time $t$ as
    \begin{equation}
    \begin{split}\label{def_total_q}
    h(t) &\triangleq \hat{h}(\boldsymbol{Q}(t))\! =\!\! \sum_{k=1}^{K} Q^p_k(t) \!+\! \alpha \sum_{s \in \mathcal{S}}\sum_{f \in \mathcal{F}_s} Q^{s}_{f}(t)
    \end{split}
    \end{equation}
    where $\alpha$ is a constant that weights the importance of stabilizing instances queues compared to prediction queues.
    Accordingly, we define the queue stability \cite{neely2010stochastic} as
    \begin{equation}\label{stability}
    \displaystyle \limsup_{T \to \infty} \frac{1}{T} \sum_{t=0}^{T-1} \mathbb{E} \left\{ h(t) \right\} < \infty.
    \end{equation}
\subsection{Problem Formulation}
Based on the above models, we formulate the following stochastic network optimization problem (\textbf{P1}) that aims at the joint minimization of time-average expectations of weighted communication cost and energy cost while ensuring queue stability. With such formulation, we explore the potential trade-off among different system metrics. 
\begin{equation}\label{problem_long_term}
	\setstretch{1.5}
	\begin{array}{ccl}
		\textbf{P1:} &
	    \underset{
	    \{\boldsymbol{\mu}(t), \boldsymbol{X}(t), \boldsymbol{Y}(t) \}_{t}
	    }{\text{Minimize}} &
	    \displaystyle \limsup_{T \to \infty} \frac{1}{T} \sum_{t=0}^{T-1} \mathbb{E} \left\{  m(t) + \gamma g(t) \right\}  \\
        & \text{ Subject to } & \displaystyle 
        {\mu}^{s}_k(t) \in \mathbb{Z}_{+}, \enskip \forall k \text{ and } s \in \mathcal{S}_{f_{k, 1}}
        \\
        & & {Y}^{s}_f(t) \in \mathcal{O}_{f}, 
        \enskip\!\! \forall s \in \mathcal{S},  
        f \in \mathcal{F}_{s}
        \\
        & & (\ref{constraint_mu}), (\ref{constraint_x}), (\ref{constraint_y}), (\ref{stability}),
	\end{array}
\end{equation}
where $\gamma \geq 0$ is a constant that weights the relative importance of energy efficiency to reducing communication cost.

\section{Algorithm Design and Performance Analysis}  \label{sec: algorithm design}
We present POSCARS, an online and predictive algorithm that solves problem \textbf{P1} through a series of online decisions,
followed by its performance analysis and three variants.

\subsection{Algorithm Design}
Problem \textbf{P1} is challenging to solve due to time-varying system dynamics, the online nature of request arrivals, and complex interaction between successive VNF instances.
Therefore, instead of solving problem \textbf{P1} directly, 
we adopt Lyapunov optimization techniques \cite{neely2010stochastic} to transform the long-term stochastic optimization problem 
into a series of sub-problems over time slots, as specified by the following lemma. 
\begin{lemma} \label{lemma: transformation}
	\textit{
		By applying Lyapunov optimization techniques and the concept of \textit{opportunistically minimizing an expectation}, problem \textbf{P1} can be transformed to the following optimization problem to be solved in each time slot $t$:
		\begin{equation}\label{problem p2}
	\setstretch{1.3}
	\begin{array}{ccl}
		\textbf{P2:} &
	    \underset{\boldsymbol{\mu}, \boldsymbol{X}, \boldsymbol{Y} }{\text{Minimize}} & 
	    \displaystyle
	    \sum_{k=1}^{K}  
					\sum_{s \in \mathcal{S}_{f_{k,1}}}
					\left[  
						- Q^{p}_{k}(t) + 
						\alpha 
						Q_{f_{k,1}}^{s}(t)
					\right]  
					\mu^{s}_{k} 
	    \\
	    & & \displaystyle
	    + 
		\sum_{f \in \mathcal{F}_{nt}}
		\sum_{s' \in \mathcal{S}_{f}}  
		\sum_{s \in \mathcal{S}_{n(f)}}
	    l^{(s', s)}_{f}(t)
	    X^{(s', s)}_{f} 
	    \\
	    & & \displaystyle
	    +
	    \sum_{s\in\mathcal{S}}
	    \sum_{f\in\mathcal{F}_{s}}
	    r^{s}_{f}(t, Y^{s}_{f})
	    \\
	 \end{array}
	 \end{equation}
	 \begin{equation}
	 \setstretch{1.3}
	\begin{array}{ccl}
        & \text{Subject to} & \displaystyle
        (\ref{constraint_mu}), (\ref{constraint_x}), (\ref{constraint_y}) 
        \text{ and }
        {\mu}^{s}_k \in \mathbb{Z}_{+}, \enskip \forall k, s \in \mathcal{S}_{f_{k, 1}}
        \\
		& & \displaystyle
        {X}^{(s', s)}_{f} \in \{0, 1\}, 
        \enskip \forall s', s \in \mathcal{S}, \enskip f \in \mathcal{F}_{s} 
        \\
        & & {Y}^{s}_{f} \in \mathcal{O}_{f}
        \enskip \forall s \in \mathcal{S}, f \in \mathcal{F}_{s}.
	\end{array}
	\end{equation}
	where $l^{(s', s)}_{f}(t)$ is defined as
	\begin{equation}\label{def_l}
		l^{(s', s)}_{f}(t) \triangleq 
		\left[
		Vw_{s', s}\left(t\right) 
		+ \alpha Q_{n(f)}^{s}\left(t\right)
		\right]
		B^{s'}_{f}(t),
	\end{equation}
	such that $V$ is a positive parameter that weights the importance of minimizing system cost compared to stabilizing system queues,
	and $r^{s}_{f}(t, Y)$ is defined as
	\begin{equation}\label{def_r}
		r^{s}_{f}(t, Y) \triangleq 
		V\gamma\boldsymbol{\lambda}_{s}^{T}Y-\alpha Q_{f}^{s}\left(t\right)\phi_{f}\left(Y\right).
	\end{equation}
	}
\end{lemma}

The detailed proof of Lemma \ref{lemma: transformation} is relegated to Appendix-A. 
Here we provide a sketch of how the problem transformation is carried out. Note that the key technique we adopt is the \textit{drift-plus-penalty} method \cite{neely2010stochastic}, which generally aims to stabilize a queueing network while also optimizing the time-average of some objective (\textit{e.g.}, the total cost of energy consumption and communication in \textbf{P1}). To this end, a quadratic function (\textit{a.k.a.} Lyapunov function) is first introduced to characterize the stability of all queues in each time slot. Then the key idea of the method is to introduce a drift-plus-penalty term to characterize the joint change in the queue stability and the objective value across time slots. In particular, the drift-plus- penalty term is defined as the weighted sum of two parts. One is defined as the difference (\textit{a.k.a.} drift) between the Lyapunov functions of two consecutive time slots, which measures the short-term change in queue stability. The other part is defined as the instant objective value in a time slot. Then the stability of the queueing network and the optimization of the time- average of the objective are jointly achieved by deriving an online control policy that greedily minimizes the upper bound of the drift-plus-penalty term during each time slot. In this way, it can be proven that it is equivalent to solve problem \textbf{P1} by resolving a series of subproblems (\textbf{P2}) over time slots.

Note that by solving problem \textbf{P2} over time slots, problem \textbf{P1} can be solved asymptotically optimally as the total number of time slots $T$ and the value of parameter $V$ both approach infinity, as shown by Theorem 1 in Sec. \ref{subsec: perf analysis}.
Furthermore, problem \textbf{P2} can be decomposed into three sub-problems for request admission, service chaining, and resource allocation, with their decisions in each time slot denoted by $\boldsymbol{\mu}$, $\boldsymbol{X}$, and $\boldsymbol{Y}$, respectively.
	Then we propose POSCARS, a predictive online service chaining and predictive resource scheduling scheme, and
	show its pseudocode in Algorithm \ref{algo}.
		\begin{remark}
			Regarding request admission,
			when all instances are more loaded than the prediction queue. 
			in order not to overload any instances, POSCARS admits
			only untreated requests at current time slot and spreads them evenly onto least loaded instances.
			However, when instances all have shorter queue lengths than the prediction queue, 
			POSCARS admits all future requests and assigns them to the least loaded instances. 
		\end{remark}
		\vspace{-0.1em}
		\begin{remark}
			POSCARS decides the 
			service chaining 
			by jointly considering instances' queue length and the communication cost.
			Recall the definition in (\ref{def_l}), where 
			the weighted summation $\alpha Q^{s}_{n(f)}(t) + V w_{s', s}(t)$ actually reflects 
			the unit price of sending a request from VNF $f$'s instance on server $s'$ to the instance of its next VNF on server $s$.
			If the target instance is heavily loaded, there will be a high price of forwarding the request to that instance.
			Besides, a large communication cost $w_{s', s}(t)$ also makes it less willing to choose the target instance.
		\end{remark}
\begin{algorithm}[!t]
 \caption{POSCARS (Predictive Online Service Chaining And Resource Scheduling) in one time slot}
 \begin{algorithmic}[1] \label{algo}
 \STATE Initially in time slot $t$, given backlog sizes $\boldsymbol{Q}(t)$, service rates $\boldsymbol{B}(t)$, energy cost $\{\boldsymbol{\lambda}_{s}\}$, and communication cost $\boldsymbol{w}(t)$. Output: chaining and scheduling decisions.
    \STATE \textbf{for} 
	every network service $k \in \{1, 2, \dots, K\}$
    \STATE $~~$ \%\% Request admission for ingress VNF $f_{k, 1}$
  	\STATE $~~$ The traffic classifier first finds the set $S^{*}_{f_{k,1}}$ of servers 
  	$~~~$ that host the least loaded instances of VNF $f_{k,1}$.
  	\\
      \STATE $~~$ \textbf{if} $\alpha Q^{s}_{f_{k,1}}(t) > Q^{p}_{k}(t)$ for all $s \in S^{*}_{f_{k,1}}$
      	\textbf{then}
  	  \STATE 
  	  $~~$ $~$
  	  Admit the $Q^{(0)}_{k}(t)$ untreated requests at current time. \\
  	  \STATE $~~$ \textbf{else} \\
  	  \STATE $~~$ $~$
  	  Admit all $Q^{p}_{k}(t)$ untreated requests. \\
  	  \STATE $~~$ \textbf{endif} \\
  	  \STATE $~~$ Spread admitted request evenly to least loaded instances. \\
  	  \STATE \textbf{endfor}
      \STATE \%\% Service chaining
  	  \STATE \textbf{for} every non-terminal VNF $f \in \mathcal{F}_{nt}$:
  	  \STATE $~~$\textbf{for} the instance of $f$ on server $s' \in \mathcal{S}_{f}$: \\
  	  \STATE $~~~~~$ Forward its processed requests to one of the servers \\ 
  	  		 $~~~~~$ from $\mathcal{S}_{n(f)}$ with
  	  minimum $l^{(s', s)}_{f}(t)$. 
  	  \\
  	  \STATE $~~$ \textbf{endfor}
  	  \STATE \textbf{endfor} \\
	  \STATE \%\% Resource scheduling
  	  \STATE \textbf{for} every server $s \in \mathcal{S}$:\\
  	  \STATE $~~$ 
  	  Initialize an empty lookup table $\mathcal{Y}_{cand}$ and set $\mathcal{F}_{alloc}$.
  	  \STATE $~~$ Set $\mathcal{Y}_{cand}[r^{s}_{f}(t, Y)] \leftarrow (f, Y),\  \forall\ f \in \mathcal{F}_{s}$ and $Y\in \mathcal{O}_{f}$ 
  	  \STATE $~~$ \textbf{while} $|\mathcal{Y}_{cand}| > 0$:
  	  \STATE $~~~~~$ Choose the minimum $r^{*}$ among all keys of $\mathcal{Y}_{cand}$. 
  	  \STATE $~~~~~$ Select its associated $f^{*}$ and $Y^{*}$.
  	  \STATE $~~~~~$ Remove entry with key $r^{*}$ from $\mathcal{Y}_{cand}$.
	  \STATE $~~~~~$ \textbf{if} $r^{*} < 0$ and 
	    	  $\sum_{f \in \mathcal{F}_{alloc}} Y^{s}_{f}(t) + Y^{*} \preceq \boldsymbol{c}_s$:
	  \STATE $~~~~~~~~$	Allocate resource to $f^{*}$ according to $Y^{*}$.
	  \STATE $~~~~~~~~$	$\mathcal{F}_{alloc} \leftarrow \mathcal{F}_{alloc} + \{f^{*}\}$.	  
	  \STATE $~~~~~~~~$	Remove all entries related to $f^{*}$.
	  \STATE $~~~~~$ \textbf{endif} 
  	  \STATE $~~$ \textbf{endwhile}
  	  \STATE \textbf{endfor} \\
\end{algorithmic}
\end{algorithm}
\setlength{\textfloatsep}{0pt}
		\begin{remark}
			On server $s$, the resource allocation is decided by jointly considering the resource cost and the queue length of its resident instances. 
			Particularly, 
			we regard the term $V \gamma \boldsymbol{\lambda}_{s} - \alpha Q^{s}_{f}(t) \boldsymbol{\phi}_{f}$ as the unit net cost vector of resources allocated to the instance of VNF $f \in \mathcal{F}_{s}$.
			Regarding the unit net cost of type-$i$ resource, \textit{i.e.},
			$V \gamma \lambda_{s, i} - \alpha Q^{s}_{f}(t) \phi_{f}$,
			it is the weighted difference between the unit cost $\lambda_{s, i}$ of type-$i$ resource and the queue length $Q^{s}_{f}(t)$ of the instance. 
			A high unit resource cost will result in a prudent allocation. 
			On the other hand, a sufficiently long queue length will make the allocation more worthwhile. 
			In both cases, POSCARS selects the set of resource allocation decisions $\{Y^{s}_{f}\}_{f \in \mathcal{F}^{s}}$ that satisfy constraint (\ref{constraint_y}) and minimize the total net cost. 
		\end{remark}

\subsection{Performance Analysis}  \label{subsec: perf analysis}
\vspace{-0.32em}
We analyze the computational complexity of POSCARS in each time slot as follows. 
For each network service, it takes $O(|\mathcal{S}|)$ time to make request admission decisions (lines $4$-$10$).  
Next, each non-terminal VNF instance selects and forwards requests to its successors in $O(|\mathcal{S}|)$ time (line $15$). 
Every server takes $O(|\mathcal{F}|)$ time to initialize the lookup table (lines $21$-$23$) and $O(\Omega_{max} \times |\mathcal{F}|)$ time to decide the resource allocation, where $\Omega_{max}$ is the maximum number of applicable resource allocation for any VNF instance. 
In practice, POSCARS can be run in a distributed manner. 
Particularly, the request admission sub-routine can be implemented on each traffic classifier with a computational complexity of $O(K \times |\mathcal{S}|)$; 
meanwhile, the service chaining and resource scheduling sub-routines can be deployed on the hypervisor of each server, with computational complexities of $O(|\mathcal{S}|)$ for each instance and $O(\Omega_{max} \times |\mathcal{F}|)$ for each server, respectively, where $\Omega_{max}$ is the maximum number of applicable resource allocations.

On the other hand, without predictive scheduling, 
we show that POSCARS achieves an $[O(V), O(1/V)]$ trade-off between the time-averages of total queue length and total cost
via the tunable parameter $V$. 
In particular, given the value of $\gamma$, let $(m^{*} + \gamma g^{*})$ denote the optimal value of problem \textbf{P1}; then we have the following theorem. 
\begin{theorem}
	\textit{
		 Suppose that $h(0)<\infty$ and, given the system resource capacities on each server and VNF placement, there exists an online scheme which ensures that, for each VNF instance, the mean arrival rate is smaller than its mean service rate. 
		Under POSCARS without prediction, there exist constants $B>0$ and $\epsilon>0$ such that
		\begin{equation*}
			\begin{array}{l}
				\displaystyle \limsup_{T \to \infty} \frac{1}{T} \sum_{t=0}^{T-1} \mathbb{E} \left\{  m(t) + \gamma g(t) \right\} \leq \frac{B}{V} + m^{*} + \gamma g^{*},
			\\
				\displaystyle \limsup_{T \to \infty} \frac{1}{T} \sum_{t=0}^{T-1} \mathbb{E} \left\{ h(t) \right\} \leq 
				\frac{B+V (m^{*} + \gamma g^{*})}{\epsilon}.
			\end{array}
		\end{equation*}
	}
\end{theorem}

The proof is relegated to Appendix-B.
{Theorem 1 demonstrates an $[O(V), O(1/V)]$ trade-off between system cost optimization and queue stability. 
Particularly, without prediction, POSCARS can achieve a near-optimal cost within an $O({1}/{V})$ optimality gap but at the cost of an $O(V)$ increase in the time-averaged total queue length. 
Intuitively, with a large value for $V$, VNF instances are more willing to steer requests to their successive instances in nearby servers, while server would allocate resources to instance with less energy cost. As a result, the total cost can be effectively reduced; however, some servers may become hot spots and the total queue length will increase.
In contrast, a smaller value of $V$ conduces to more balanced queue loads among servers and more energy cost consumed to serve requests, leading to an increasing total cost.
Moreover, given predicted information about future requests, POSCARS can achieve a better trade-off with a notable delay reduction by pre-serving requests with surplus system resources.
We verify such advantages by our simulation results in Section \ref{sec: simulation}. 
}


\subsection{Practical Issues and Variants of POSCARS}

The distributed nature of POSCARS requires each VNF instance to gather relevant system dynamics on its own.
However, the probing process may incur considerable sampling overheads and additional latencies. 
Meanwhile, each instance makes its independent decision based on the sampled information 
at the beginning of a time slot.
Therefore, instances may blindly choose the same lowest-cost instance, without knowing others' choices. 
The chosen instance will then become overloaded due to the non-coordinated decisions. 
An alternative is to perform sampling before sending each request. 
Nonetheless, this method suffers from the messaging overheads of frequent samplings. 
A possible compromise is to split the processed requests into batches, then sample and schedule for each batch separately.

To mitigate such issues, we propose the following variants of POSCARS, by adopting the ideas from recent randomized load balancing techniques,
such as \textbf{The-Power-of-$d$-Choices} \cite{mitzenmacher2001power}, \textbf{Batch-Sampling} \cite{ousterhout2013sparrow}, and \textbf{Batch-Filling} \cite{ying2015power}. 

\textbf{POSCARS with The-Power-of-$d$-Choices (P-Po$d$):}
To reduce sampling overheads, we apply the idea of \textit{The-Power-of-$d$-Choices} to POSCARS. 
Particularly, every non-terminal instance probes
only the $d$ instances uniformly randomly from its next VNF. 
Next, the instance chooses to send all its processed requests to the lowest-cost instance among the $d$ samples. 
In such a way, each instance requires only few times of sampling to decide its target instance. 
Although the selected instance may not be the least-cost one,
our later simulation results show that the reduced sampling brings only a mild increase in the total cost.

The above variant significantly reduces the sampling overheads. However, the issue of non-coordinated decision making still remains.
To mitigate such issues, we adopt the idea of \textit{batch-sampling}\cite{ousterhout2013sparrow} 
 and \textit{batch-filling} \cite{ying2015power} and propose another two variants of POSCARS, namely \textit{POSCARS with Batch-Sampling} (P-BS) and \textit{POSCARS with Batch-Filling} (P-BF), respectively.
Basically, these two variants split the processed requests on each instance into batches, each batch with a size of $b$, 
then carry out scheduling upon such request batches. 
When $b\!=\!1$, we actually perform scheduling for each request separately.
When $b$ is greater than the number of processed requests, then scheduling is only performed once in a time slot, degenerating to POSCARS.
We elaborate the design of P-BS and P-BF as follows.

\textbf{POSCARS with Batch-Sampling (P-BS):}
Given an instance with $z$ batches of requests, it probes $d_{\text{bs}}z$ instances uniformly randomly from its next VNF,
where $d_{\text{bs}}$ is the respective probe ratio. 
Then the instance sends the $z$ request batch to the least-cost $z$ instances, with each batch to a distinct target instance.

\textbf{POSCARS with Batch-Filling (P-BF):}
Given an instance with $z$ request batches, it probes $d_{\text{bf}}z$ instances uniformly randomly from its next VNF. 
Then it forwards the request batches one by one. Each batch is sent to the least-cost instance among the $d_{\text{bf}}z$ samples. 
The chosen instance's cost is updated after it receives the batch of requests.

\section{Simulation}  \label{sec: simulation}
We conduct trace-driven simulations to 
evaluate the performance of POSCARS and its variants. The request arrival measurements are drawn from real-world systems\cite{benson2010network}, with a mean arrival rate of $25.5$ per time slot ($10$ms) and mean inter-arrival time of $0.594$ms. 
Besides, we conduct simulations with the Poisson request arrivals with the same rate of $25.5$. 
All the results are obtained by averaging measurements collected from $50$ repeated and independent simulations.

\subsection{Simulation Settings}

\textbf{Substrate Network Topology:}
We construct the substrate network based on two widely adopted topologies,
\textit{i.e.}, Jellyfish\cite{jellyfish} and Fat-Tree\cite{fat-tree}.
Both topologies have a comparable scale to clusters in data center networks, 
each equipped with $720$ switches, $24$ servers with deployed VNFs, and the rest $3456$ servers as hosts that generate service requests.  
Particularly, in Fat-Tree, there are $24$ pods, each pod containing to $144$ servers; amongst them, we choose one server uniformly at random as the one with deployed VNFs and the rest as hosts. 
Requests can be processed on servers in any pod with the VNF they demand.
Between any two servers, request traffic traverses over the shortest path with link capacity of $40$Gbps. 
For each pair of servers, the communication cost per request is proportional to the number of hops of the shortest path between them, with $10\%$ variation.

{\textbf{Server Resources:}}
We consider CPU cores as the resources on each server, since CPUs have become the major bottleneck for request processing in NFV systems \cite{mehraghdam2014specifying,addis2015virtual,callegati2015dynamic}. 
Servers are heterogeneous, each with a number of CPU cores ranging from $16$ to $64$.
In every time slot, we calculate the power consumption in the unit of utilized CPU cores, with $\lambda_{s} \in [1, 3]$.
Regarding parameter $\gamma$, setting it with a greater value would encourage each server to assign most resources to heavily loaded VNF instances. 
Conversely, a smaller value of $\gamma$ would lead to more balanced resource allocation among such instances; consequently, this will minimize the impact of imbalanced queue loads on the decision making for service chaining. 
The value setting depends on the objectives to fulfill in real systems. 
In our simulation, by fixing $\gamma = 1$, we assume that communication cost reduction and system energy efficiency are equally important.
  
{\textbf{Service Function Chains:}}
We deploy five network services, each with a service chain length varying from $3$ to $5$.
Each service contains at least one of the most commonly-deployed VNFs; 
\textit{e.g.}, Intrusion Detection System (IDS), Firewall (FW), Load Balancer (LB).
The rest VNFs of each service are chosen uniformly from other $30$ commonly-used VNFs \cite{li2015software} at random without replacement. 
For each VNF, the total number of instances ranges from $12$ to $18$. 

{\textbf{Prediction Settings:}}
Network services' traffic often varies in predictability. We denote the average window size by $D$, and set each service window size by sampling uniformly from $[0, 2\!\times\!D]$ at random.
We evaluate the cases with perfect and imperfect prediction. 
For perfect prediction, future request arrivals in the time window are assumed perfectly known to the system and can be pre-served. 
In practice, such an assumption is not feasible for stateful requests; nonetheless, that can be seen as the extended case of our results with more constraints on request processing. 
For imperfect prediction, the failure of prediction generally falls into two categories. 
One is \textit{false-negative} detection,
\textit{i.e.}, a request is not predicted to arrive, and as a result, it receives no pre-service before its arrival. 
The other is \textit{false-positive} detection, \textit{i.e.}, a request that does not exist is predicted to arrive. In this case, the system pre-allocates resources to pre-serve such requests. 
We consider two extreme cases: one is that we fail to predict the arrivals of all future requests; 
the other is that we correctly predict the actual future arrivals, and furthermore, some extra arrivals are falsely alarmed.
Note that any form of mis-prediction can be seen as a superposition of such two extremes. 
In addition, we also implement five schemes that forecast request arrivals in the next time slot (with window size $D=1$), including: 
1) Kalman filter (Kalman)\cite{chui2017kalman}; 
2) distribution estimator (Distr), which generates the next estimate by independent sampling from the distribution of arrivals learned from historical data;
3) Prophet (FB) \cite{taylor2018forecasting}, Facebook's time-series forecasting procedure;
4) moving average (MA) and 5) exponentially weighted moving average (EWMA)\cite{box2015time}. 

{\textbf{Baseline Schemes:}}
We compare POSCARS with three baseline schemes, including \textit{Random}, \textit{JSQ} (Join-the-Shortest-Queue), and state-of-the-art \textit{OneHop-SCH} (OneHop scheduling)\cite{wang2016joint}. 
These schemes differ in the service chaining strategy from POSCARS. 
In Random scheme, each instance uniformly randomly sends requests to one of its successors.
In JSQ scheme, each instance sends requests to its least-loaded successor. 
In OneHop-SCH, each instances sends requests to its successor with the least communication cost and idle capacity.

{\textbf{Variants of POSCARS:}}
To compare the performance of POSCARS and its variants, we evaluate them under different settings. 
For each of the variants, we vary their probe ratio ($d$ for P-Po$d$, $d_{\text{bs}}$ for P-BS, and $d_{\text{bf}}$ for P-BF) from $2$ to $5$, and fix the batch size for P-BS and P-BF as $5$ requests per batch.
We omit the cases when the ratio is $1$ and greater than $5$. Notice that the former corresponds to the random scheme and actually leverages no load information; the latter leads to excessively fined-grained control since it induces too much sampling overheads.

{\textbf{Request Response Time Metric:}}
To evaluate the impact of predictive scheduling, 
we define a request's response time as the number of time slots from its actual arrival to its eventual completion.
If a request is pre-served before it arrives, 
then the system is assumed to respond to the request upon its arrival, and the request will experience a zero response time.

\subsection{Performance Evaluation under Perfect Prediction}

{Intuitively, POSCARS is promising to shorten the requests' response time 
by exploiting predicted information and pre-allocating idle system resources to pre-serve future requests. 
Therefore, the essential benefits of predictive scheduling come from the load balancing in the temporal dimension.}
To verify such intuition, we first consider the case with perfectly predicted arrivals, and evaluate POSCARS with ($D > 0$) and without ($D\!=\!0$) prediction, against the baseline schemes. 

\begin{figure}
  \centering 
  \includegraphics[scale=.26]{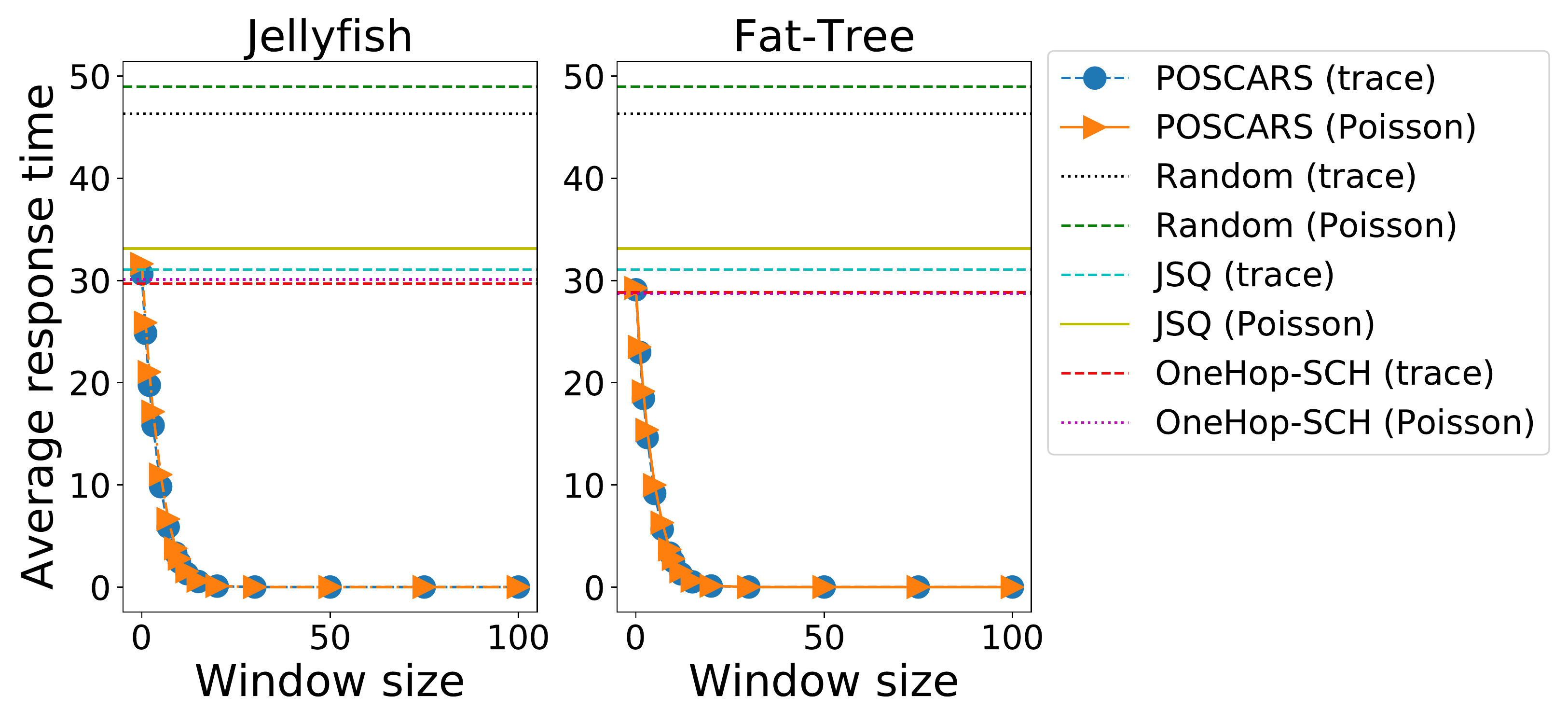}
\vspace*{-0.4cm}
\caption{Average response time ($ms$) with various window sizes given trace and Poisson arrival process, under different topologies.}
\label{win-delay-perfect}
\end{figure}

\begin{figure}[!t]
  \centering
  \subfigure[Total cost with parameter $V$]{
    \includegraphics[scale=.18]{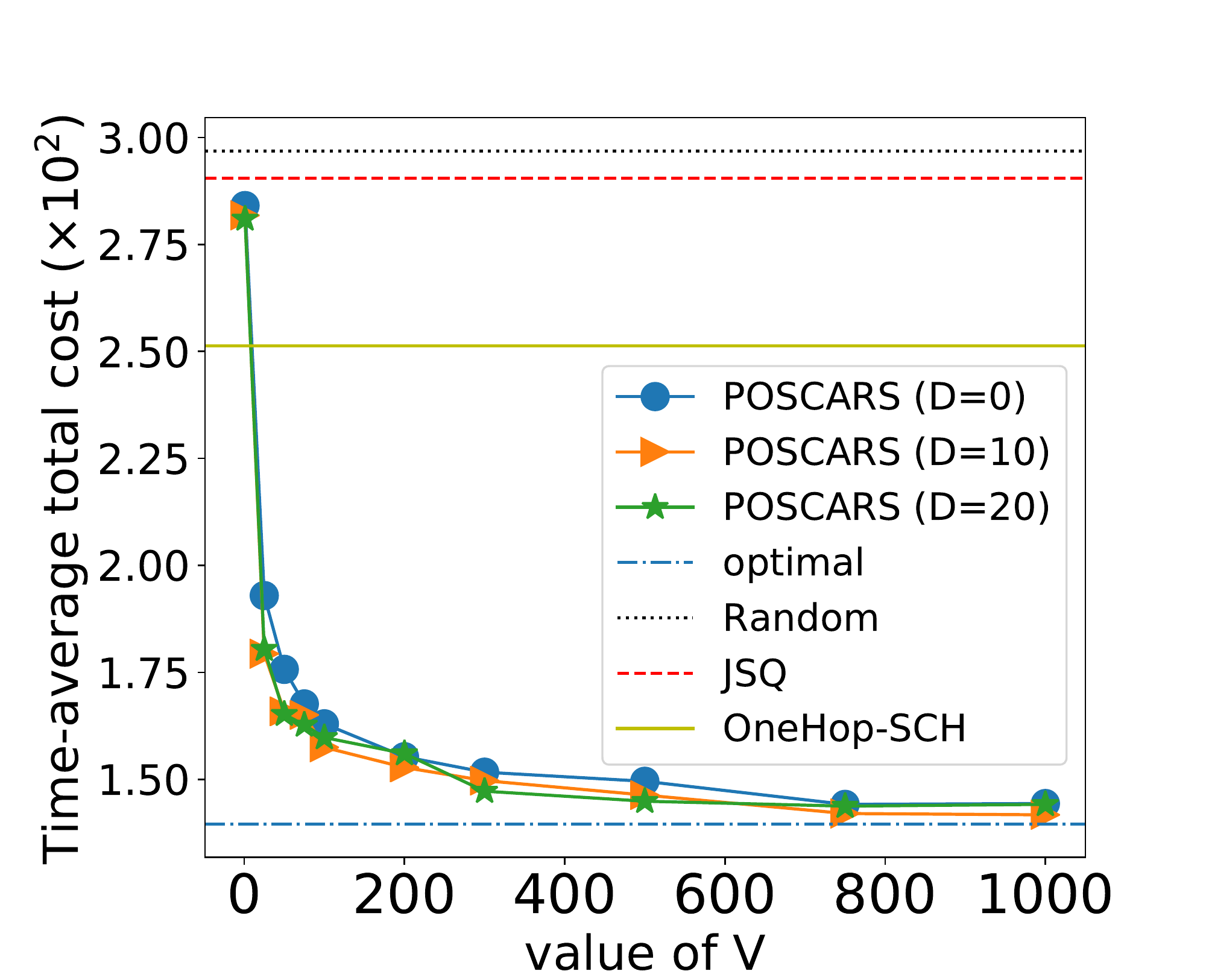}  
  }
  \hspace{-0.1cm}
  \subfigure[Queue size with parameter $V$]{ 
    \includegraphics[scale=.145]{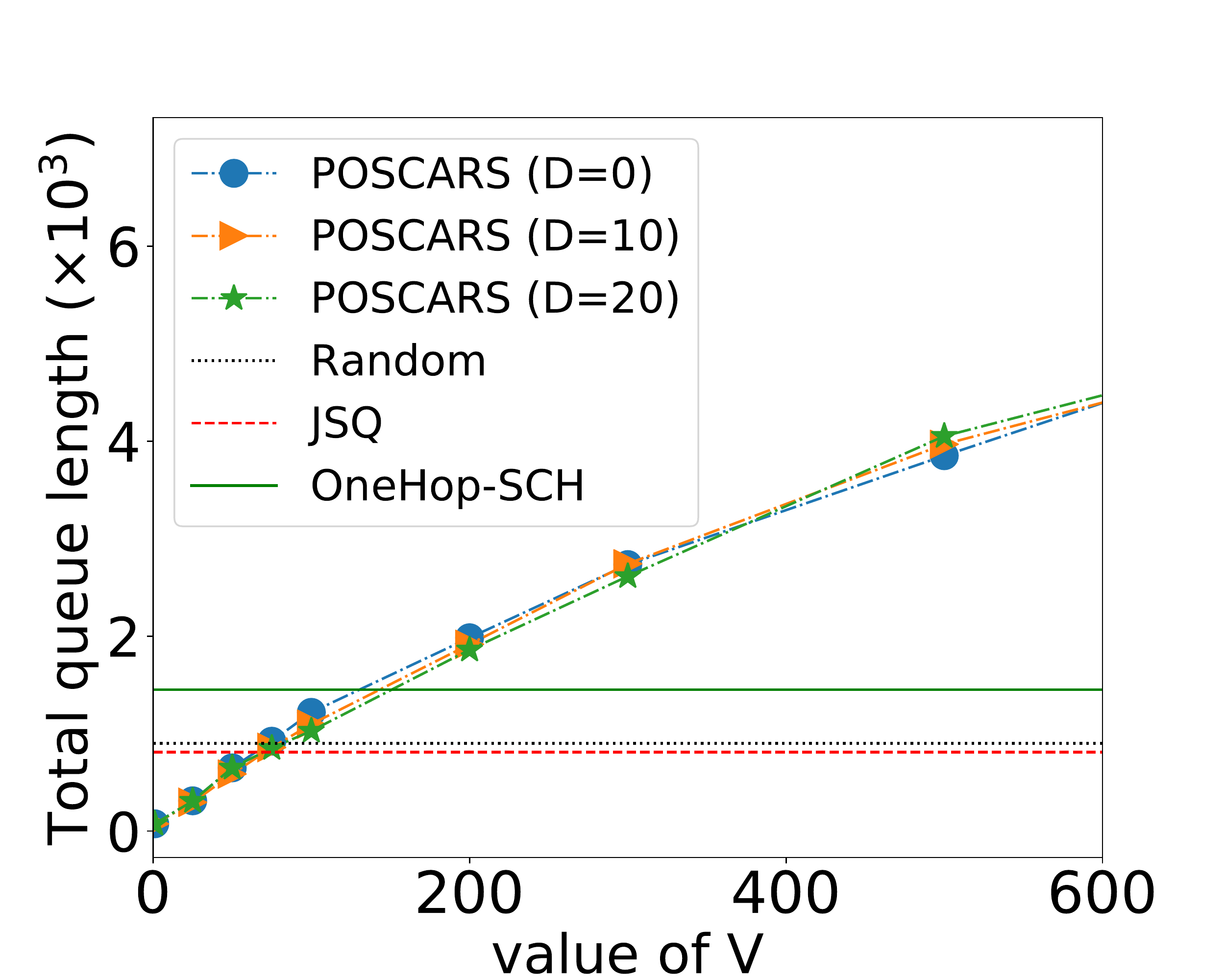}
  } 
  \caption{Total queue length under different window sizes.}
  \label{qbacklogs_comp}
\end{figure}

\textbf{Average response time vs. window size $D$:}
Figure \ref{win-delay-perfect} shows the performance of the different schemes under Jellyfish and Fat-Tree topology.
The response times induced by the baseline schemes remain constant since they do not involve predictive scheduling. 
Random incurs the highest response time ($\sim 47$ms), since it disregards information about workloads or communication cost when dispatching requests. JSQ does much better ($\sim 32$ms) because requests are always greedily forwarded to the least-loaded successors. 
OneHop-SCH outperforms the previous two by jointly taking the workloads and communication cost into consideration.
Meanwhile, without prediction ($D=0$), POSCARS achieves comparable performance with OneHop-SCH; 
but as $D$ increases from $0$ to $20$, we observe a significant reduction in the average response time under both topologies;
\textit{e.g.}, from $29.1$ms to $0.5$ms under Fat-Tree topology. 
The marginal reduction diminishes as $D$ further increases, and eventually, remains at around $0.2$ms.  

\textit{Insight:}
In practice, due to traffic variability, it is often not realistic to achieve high predictability (large $D$). However, the results show that, 
only mild-value of future information suffices to POSCARS's shortening requests response time effectively and achieving load-balancing in the temporal dimension. 
With more future information, the reduction diminishes since the idle system resources have already been depleted.

Considering the qualitative similarities among curves with different settings, we only present results under Fat-Tree and trace-driven request loads.

\textbf{Backlog-cost trade-off with parameter $V$:}
Recall from Section III.B that the value of parameter $V$ controls the backlog-cost trade-off. Figures \ref{qbacklogs_comp}(a) and \ref{qbacklogs_comp}(b) verify such a trade-off. 
Figure \ref{qbacklogs_comp}(a) compares the time-average communication cost of POSCARS with $D = 0$, $10$, $20$, against baselines. 
Both Random and JSQ induce a high total cost since their decision making disregards the resultant communication cost and the heterogeneity of servers in terms of energy cost. 
OneHop-SCH further lowers the total cost by about $13.6\%$, by taking its advantages of jointly optimizing cost and shortening queue lengths based on flow-level statistics. 
Given different choices of $D$, POSCARS achieves close-to-optimal time-average total cost as the value of $V$ rises up to $10^3$. Notably, POSCARS excels OneHop-SCH whenever $V > 10$.

However, recall that parameter $V$ weighs the importance of minimizing system cost compared to maintaining queue stability. 
Hence, to reduce system cost, large values of $V$ also lead to increased backlogs. 
By \textit{Little's theorem}\cite{little1961proof}, this would increase response time as well. In Figure \ref{qbacklogs_comp}(b), we see that the total queue length is almost proportional to value of $V$, exceeding all other baselines as $V > 150$. 

\textit{Insight:}
POSCARS achieves a backlog-cost trade-off with different values of parameter $V$. By choosing an appropriate value of $V$ from $[10, 150]$, it outperforms the baseline schemes with both lower system cost and shorter queue lengths. 
In practice, such an interval may vary from system to system but it is usually proportional to the ratio of magnitudes of the total queue length to total system cost.

\textbf{POSCARS and its variants:}
Upon forwarding requests, POSCARS requires each instance to collect statistics from all its successors. 
In practice, this may require non-negligible sampling overheads in face of a large number of instances. 
In Section III.C, we propose three variants of POSCARS, \textit{i.e.}, P-Po$d$, P-BS, and P-BF. These variants trade off optimality of decision making for reduction in sampling overheads and complexity\cite{ying2015power} from $O(n)$ to $O(1)$, where $n$ denotes the total number of candidate instances. 
Figure \ref{variants} evaluates the total cost and average response time induced by POSCARS and its variants, with parameter $V\!=\!10$, $D\!=\!1$, batch size of $5$ for P-BS and P-BF, and the probe ratio $d = d_{\text{bs}} = d_{\text{bf}} \in \{2, 5, 8\}$.

In Figure \ref{variants}(a), we see that POSCARS achieves the lowest total cost, since each instance's decision making is based on the full dynamics of its succeeding instances. 
For each variant, we see a cut-down in the total cost by up to $22.1\%$ as $d$ increases from $2$ to $8$. 
Similarly, from Figure \ref{variants}(b), we also observe a reduction in response time from about $34.3$ms by up to $17.6\%$. 
Among the three variants, P-BS and P-BF induce more reduction in both cost and response time than P-Po$d$, because aggregated sampling is often more conducive to lowering the cost \cite{ousterhout2013sparrow}.

\textit{Insight:}
By sampling partial system dynamics for decision making, variants of POSCARS trade off optimality for reduction in sampling overheads and complexity. 
Owing to aggregated sampling, P-BF and P-BS outperforms P-Po$d$ in terms of both lower total cost and response time. 

\begin{figure}
  \centering 
  \subfigure[Time-average total cost]{ 
    \includegraphics[scale=.29]{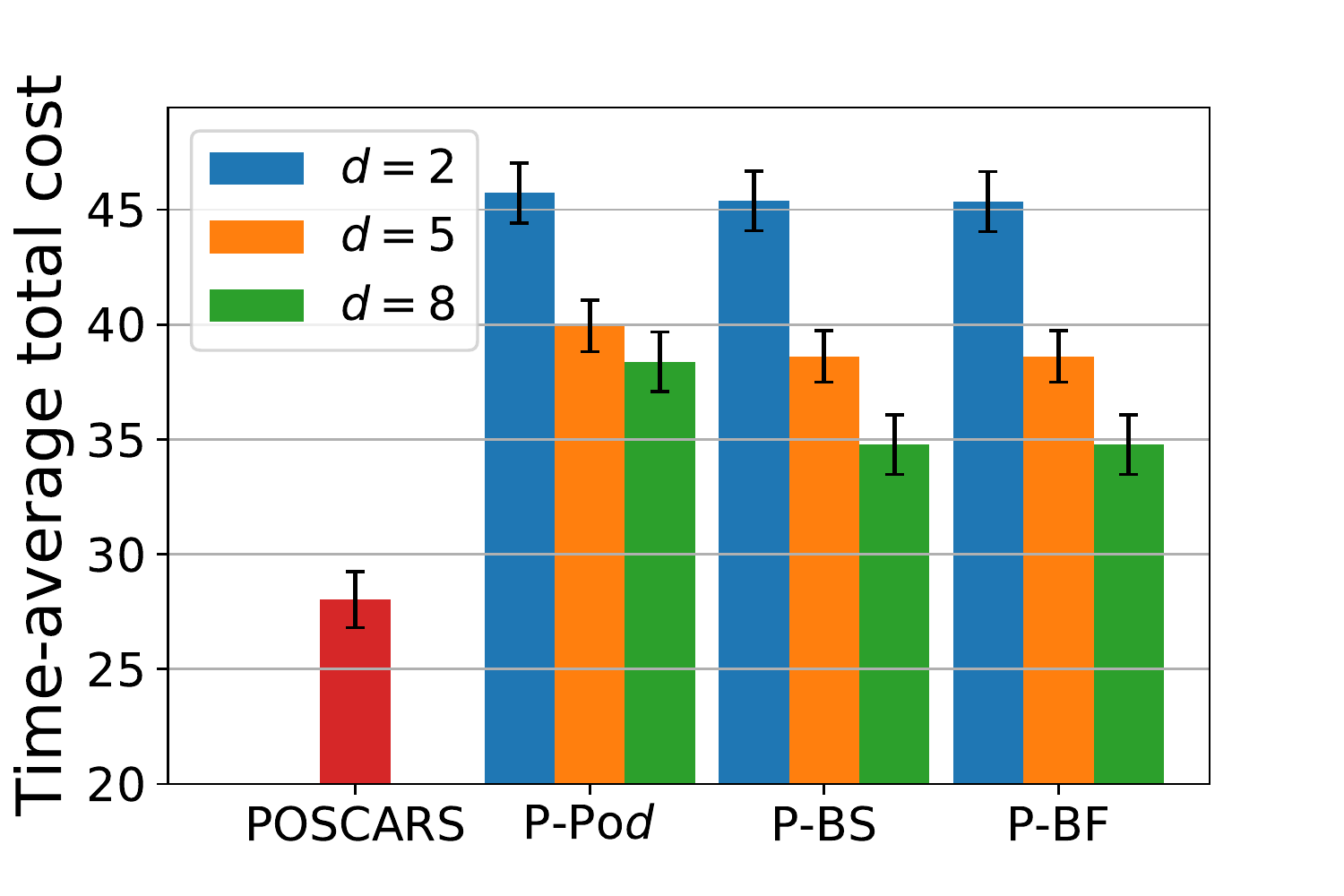}
  } 
  \hspace{-0.9cm}
    \subfigure[Average response time]{ 
    \includegraphics[scale=.29]{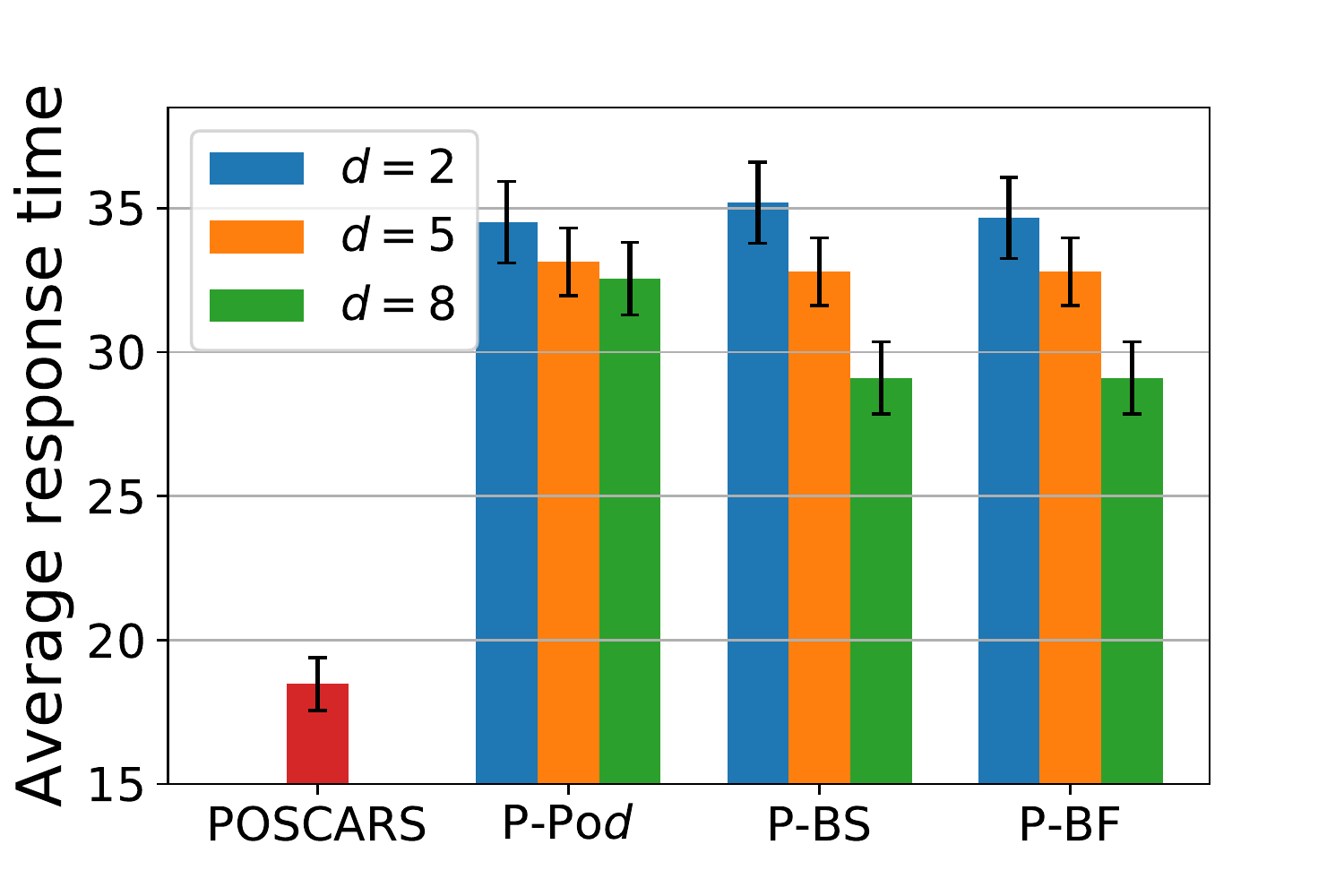}
  } 
\caption{Comparison among POSCARS and its variants}
\vspace{0cm}
\label{variants}
\end{figure}

\subsection{Performance Evaluation under Imperfect Prediction}
In practice, prediction errors are inevitable due to dataset bias and noise. To explore the fundamental limits of predictive scheduling, we evaluate the impact of imperfect prediction on the system performance.

\textbf{Total cost and response time vs. $V$:}
Figure \ref{imperfect} compares the time-average total cost and average response time induced by different forecasting schemes and perfect scheduling using POSCARS.
In Figure \ref{imperfect}(a), we observe that all forecasting schemes incur higher time-average total cost than predictive scheduling by up to $35.2\%$. 
The reason is as follows. Recall that the prediction under these forecasting schemes are imperfect, with both false-negative and false-positive detection. 
Particularly, the system pre-allocates extra resources to pre-serve false-positive requests, resulting in higher total cost. 
Figure \ref{imperfect}(b), shows the overall ascending trend proportional to increased $V$. This is due to that larger values of $V$ lead to a greater total queue length, and by \textit{Little's theorem}\cite{little1961proof}, a greater queue length implies longer response time. 
However, we also see that, even under imperfect prediction, predictive scheduling does not necessarily lead to longer response time than that under perfect prediction. 

To figure out the reason, we consider two extreme cases. 
One is all-\textit{false-negative}, \textit{i.e.}, during each time slot, all future request arrivals in the lookahead window are false-negative.
Notice that this case is equivalent to the case without predictive scheduling ($D=0$), 
since no requests will be pre-allocated resources.
The other is all-\textit{false-positive}, \textit{i.e.}, all future request arrivals are perfectly predicted, and besides, some extra requests are wrongly predicted to arrive.

\textbf{Perfect prediction with two extremes:} Figure \ref{extremes}(a) compares average response times under perfect prediction and the two extremes, with $D\!=\!\!5$, $\alpha\!=\!\!10$, and $5$ false-positive requests on average. 
Overall, the average response time is proportional to the value of $V$. Miss detection incurs higher response time than the other two, because it does not pre-serve any requests before they arrive. 
On the other hand, perfect prediction and false alarm do not necessarily outperform the other with lower response times. 
This is because of two consequences of false alarm. The \textit{first} is that false-positive requests will consume extra system resources and prolong the request queues length, thus leading to longer response times. 
The \textit{second} is that, 
according to lines $5$ - $9$ in Algorithm \ref{algo}, false-positive requests result in a greater prediction queue length. That forces POSCARS to admit future requests more frequently, thus conducing to shorter response times. The same effect can be achieved by tuning the parameter $\alpha$ -- greater values of $\alpha$ lead to less frequent admission.

How do these two consequences interplay? The question is answered by Figure \ref{extremes}(b), where the number of average false-positive requests varies from $0$ to $100$, with $D=5$, $\alpha=10$, and $V\!\!=\!\!50$.
When the average number of false-positive requests increases from $0$ to $5$, the resultant response time falls even lower than that under perfect prediction. 
In such cases, the second consequence dominates -- mild false alarm leads to more frequent admission, making POSCARS spread requests more evenly among instances. 
However, as false alarm continues aggravating, the reduction diminishes and the response time grows constantly. 
In such cases, though the admission frequency is intensified, too much false alarm severely extends the total queue length, 
offsetting and eventually outweighing the effect of load balancing.

\begin{figure}[!t]
    \centering
    \subfigure[Time-average total cost]{
    \includegraphics[scale=.18]{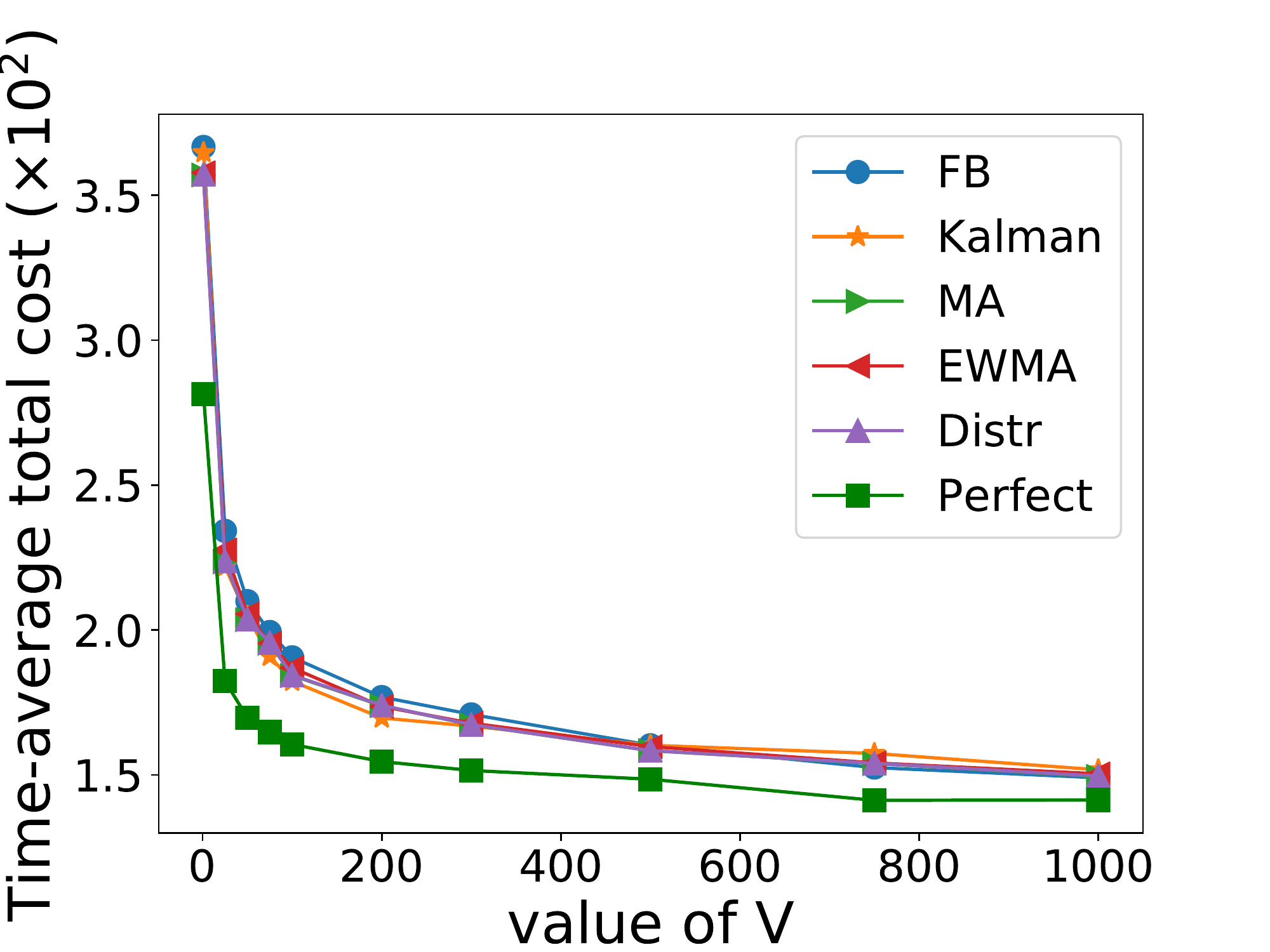}
    }
    \subfigure[Average response time]{
    \includegraphics[scale=.18]{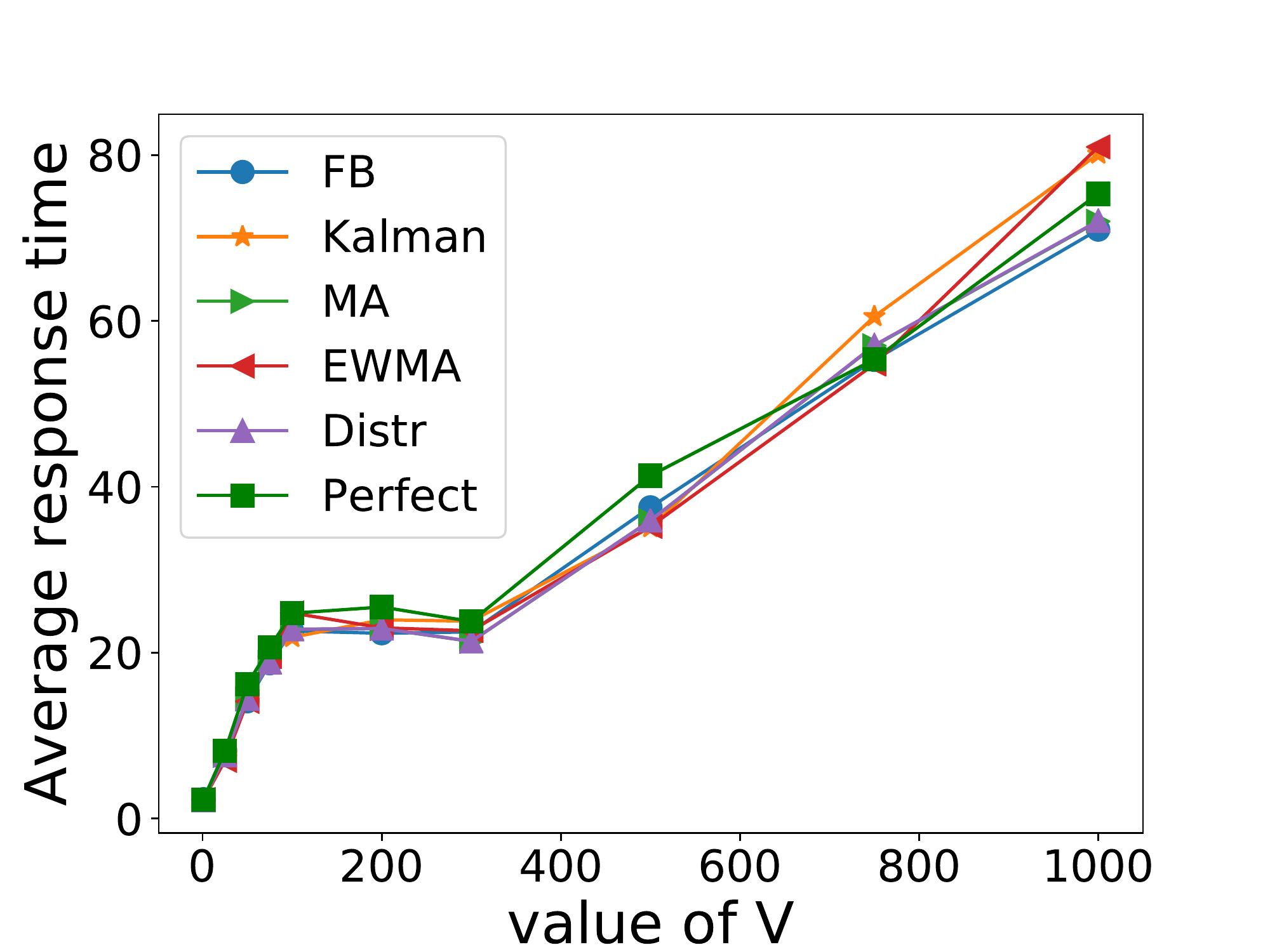}
    }
\caption{Performance of prediction schemes with $D=1$ and $\alpha=10$.}
\label{imperfect}
\end{figure}
\setlength{\textfloatsep}{0pt}
\begin{figure}[!t]
    \centering
    \subfigure[Different values of $V$]{
    \includegraphics[scale=.05]{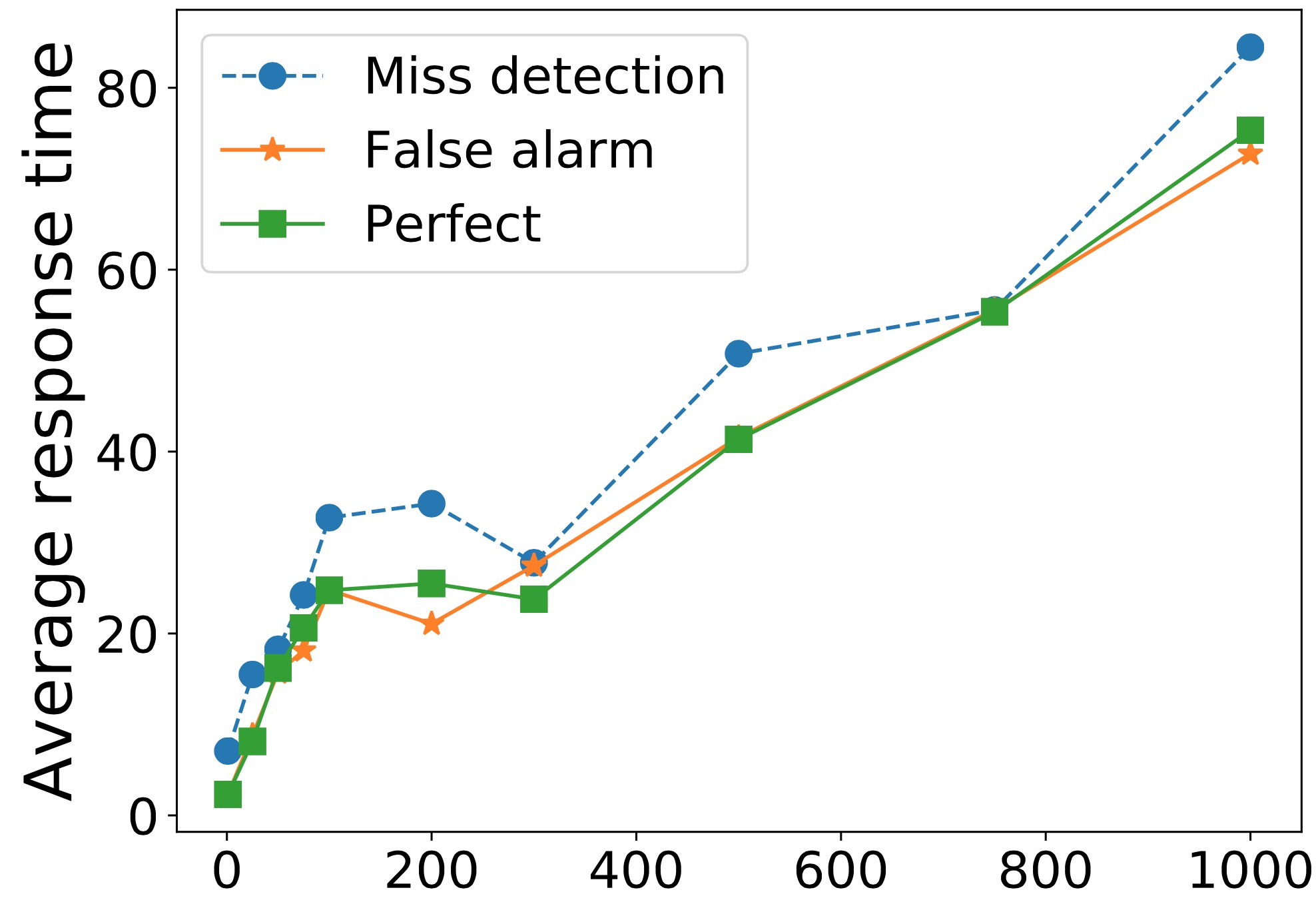}
    }
    \subfigure[Different false-alarms]{
    \includegraphics[scale=.14]{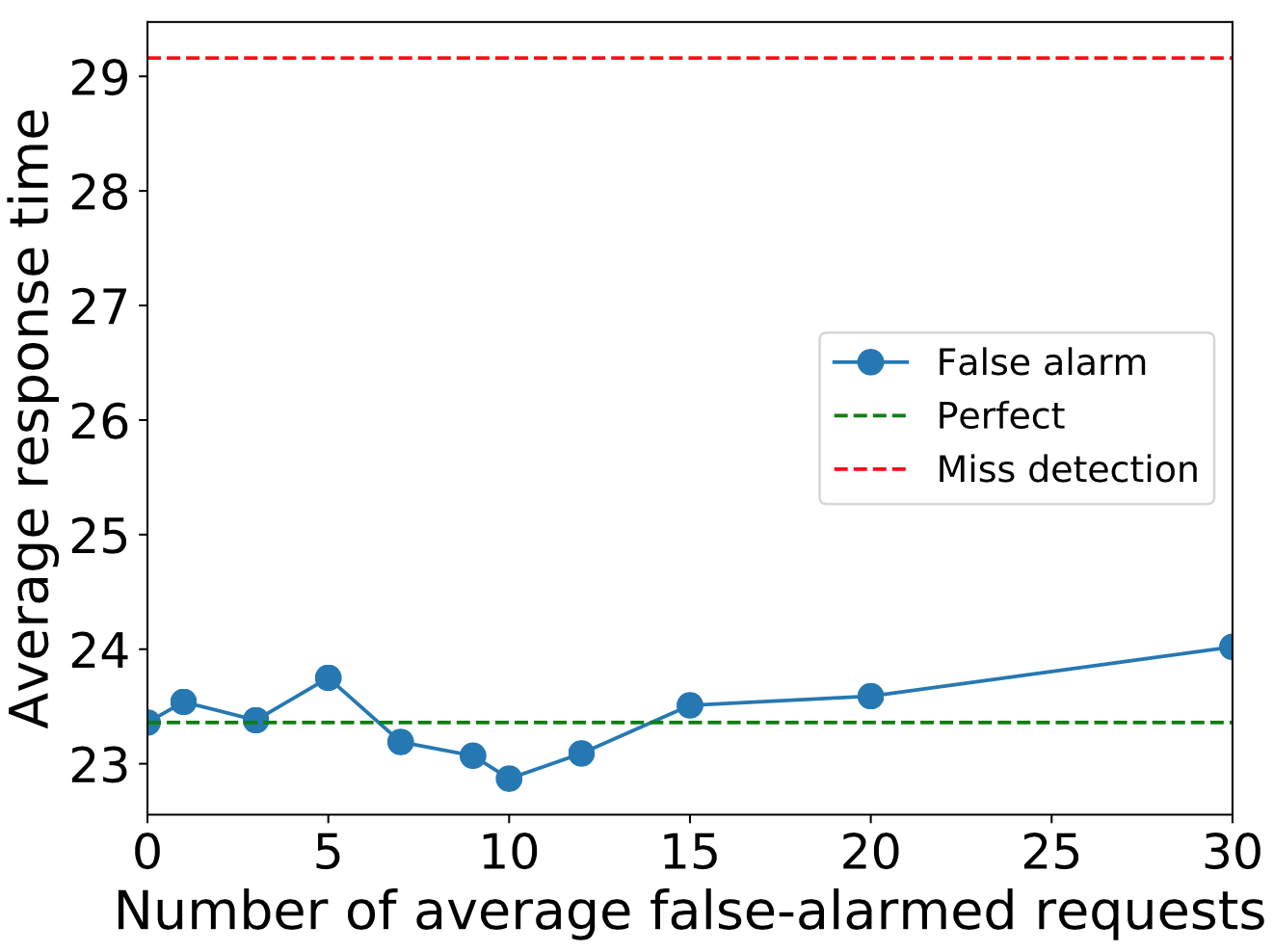}
    }
\caption{Average response time under three different prediction cases}
\label{extremes}
\end{figure}

\textit{Insight:}
Imperfect prediction does not necessarily degrade system performance, such as longer response times. Instead, mild false alarm allows the system to make better use of idle system resources, further shortening response time.

\section{Related Work}  \label{sec: related work}
In this section, we first summarize existing works that study the optimization of NFV from different aspects. Then we narrow down our focus onto those that are most relevant to this paper and compare their proposed approaches with ours.

{
\subsection{Optimizing NFV/VNF from Different Aspects }
A wide range of recent works have studied NFV systems from various aspects.
Below we take a brief overview and discuss how they are related to our work.
\begin{itemize}
	\item[$\diamond$] \textit{VNF placement}: In NFV, the placement of VNF instances often has a significant impact on system performances \cite{laghrissi2018survey} and thus deserves an elaborate design. A number of existing works have been conducted to this end (\textit{e.g.}, \cite{pham2017traffic}\cite{fei2017towards}\cite{cziva2018dynamic}\cite{agarwal2018joint}\cite{zhang2019adaptive}). In practice, such approaches can serve to decide the VNF placement, upon which our schemes can carry out their scheduling procedures accordingly.
	\item[$\diamond$] \textit{VNF Resource Allocation}: Another series of works (\textit{e.g.}, \cite{katsikas2018metron}\cite{li2018dhl}\cite{zhang2019hybridsfc}) focused on the optimization of resource allocation for VNF/NFV, with the aim to minimize VNF execution overheads and accelerate the processing speed of VNF instances. They concentrated on achieving such improvements with particular hardware designs. Different from such works, we mainly focus on exploiting predicted information to perform effective scheduling on existing NFV systems. Nonetheless, our schemes can be applied to systems built with their solutions.
	\item[$\diamond$] \textit{Load Balancing}: Existing works (\textit{e.g.}, \cite{wang2016load}\cite{wang2017multi}\cite{you2019efficient}) also developed various schemes to balance the workloads among chained VNF instances to improve resource utilization and fault tolerance while shortening delays in NFV systems. In practice, existing solutions can serve as reference points for system designers to tune the proper value of parameter $V$ for desired performance metrics. 
	\item[$\diamond$] \textit{Performance Characterization:} Another line of works have devoted their efforts to characterizing various dynamics of NFV systems such as performance interferences among VNF instances\cite{zeng2018demystifying}\cite{savi2019impact}. Insights from such works can be combined with our schemes to achieve even better performance.
\end{itemize}}

\subsection{Chaining and Resource Scheduling of VNFs in NFV:}
Regarding the optimization of VNF service chaining and resource scheduling in NFV, existing works generally fall into two categories.
 
Of the first category are the schemes that perform service chaining and resource scheduling in an offline fashion.
Typically, they assume the full availability of information about all service requests or flows. 
Based on flow abstraction, Zhang \textit{et al.} \cite{zhang2018placement} consider the joint optimization for VNF placement and service chaining.
They formulate the problem as an ILP problem and develop an efficient rounding-based approximation algorithm with performance guarantee. 
Yoon \emph{et al.} \cite{yoon2016nfv} adopt the BCMP queueing model for VNF service chains and propose heuristics to approximately minimize the expected waiting time of service chains. 
Wang \textit{et al.} \cite{wang2016joint} consider the joint optimization of service chaining and resource allocation and develop a greedy scheme that aims to place instances and schedule traffic with minimum link cost, CAPEX, and OPEX. 
Later,  D'Oro \textit{et al.} \cite{d2017exploiting} study service chaining problem from the perspective of congestion games. By formulating the problem as an atomic weighted congestion game,
they propose a distributed algorithm that provably converges to the Nash equilibrium. 
On the other hand, 
Zhang \emph{et al.} \cite{zhang2017joint} formulate a request-level optimization problem based on steady-state metrics and propose a heuristic scheme by applying techniques from open Jackson queueing network. 
However, there is no empirical evidence to show that service request arrivals follow Poisson process in NFV systems.
Different from existing works, our model and problem formulation assume no prior knowledge about underlying request traffic. 
Moreover, instead of offline or even centralized decision making, our solution is capable to perform near-optimal service chaining and scheduling
in a computationally efficient and decentralized manner. 

Of the second category are the online schemes that process requests upon their arrivals. 
Under this setting, Mohammadkhan \textit{et al.} \cite{mohammadkhan2015virtual} formulate the VNF placement for service chaining as a MILP problem based on flow abstraction and develop a heuristic to solve the problem incrementally. 
Lukovszki \textit{et al.} \cite{lukovszki2015online} develop an online algorithm that performs request admission and service chaining with a
logarithmic competitive ratio. 
Zhang \textit{et al.} in \cite{zhang2017proactive} propose a novel VNF brokerage service model and online algorithms to predict traffic demands, purchase VMs and deploy VNFs. 
Further, Fei \textit{et al.} \cite{fei2018adaptive}
develop an effective algorithm that 
performs online VNF scheduling and flow routing with predicted flow demand, 
so as to minimize the impact of inaccurate prediction and the cost of over-provisioned resources.
Later, Xiao \textit{et al.} \cite{xiao2019nfvdeep} propose an adaptive service chaining deployment scheme based on deep reinforcement learning techniques, which conducts service chaining to serve incoming requests in an online fashion.
Such schemes either resort to flow-level system dynamics and predicted information for decision making, or perform finer-grained control at the request level to optimize dedicated objectives. Our model considers such trade-offs and separates the granularity of system state and decision making. Besides, we also explore the fundamental benefits and limits of predictive scheduling, which still remains open in NFV systems.

\section{Conclusion}  \label{sec: conclusion}
In this paper, we studied the problem of dynamic service chaining and resource scheduling and systematically investigated the benefits of predictive scheduling in NFV systems. 
We developed a novel queue model that accurately characterizes the system dynamics. Then we formulated a stochastic network optimization problem and then proposed POSCARS, an efficient and decentralized algorithm that performs service chaining and scheduling through a series of online and predictive decisions. 
Theoretical analysis and trace-driven simulations showed the effectiveness and robustness of POSCARS and its variants in achieving near-optimal system cost while effectively shortening average response time.
Our results also show that prediction with mild false-positive conduces to shorter response times.
In addition, note that fair-share of resources and performance isolation among VNF instances are the key to maintaining high quality of service. Therefore, it is an interesting direction for future work to establish a more effective joint service chaining and scheduling scheme with multi-resource fairness consideration among VNF instances. Moreover, it would also be intriguing to explore the interplay between resource fairness and other performance metrics.





\section*{Appendix-A\\
	Proof of Lemma 1}
	
To solve problem (\ref{problem_long_term}), we adopt the Lyapunov optimization technique \cite{neely2010stochastic}. 
We define the quadratic Lyapunov function as
\begin{equation}\label{def-Lfunc}
    \begin{split}
	\displaystyle & 
	L(\boldsymbol{Q}(t)) \triangleq \frac{1}{2}
	\Big[ 
		\sum_{k=1}^{K} \left(Q_k^p(t)\right)^2 + 
		\alpha
		\sum_{s\in\mathcal{S}} 
		\sum_{f \in \mathcal{F}_{s}} 
		\left(Q_{f}^s(t)\right)^2
	\Big],
	\end{split}
\end{equation}
and the Lyapunov drift for two consecutive time slots as
\begin{equation}
	\begin{split}
		\displaystyle \Delta (\boldsymbol{Q}(t))  \triangleq &~ \mathbb{E} \Big\{ L(\boldsymbol{Q}(t+1)) - L(\boldsymbol{Q}(t)) \Big\vert \boldsymbol{Q}(t)\Big\},
	\end{split}
\end{equation}
which measures the conditional expected successive change in queues' congestion state.
To avoid overloading any queues in the system, it is desirable to make the difference as low as possible. 
However, striving for a short total queue length may incur considerable communication cost and computation cost. 
To jointly consider both queueing stability and the consequent system cost, we define the drift-plus-penalty function as
\begin{equation}
	\begin{split}\label{drift-plus-penalty}
		\displaystyle \Delta_V (\boldsymbol{Q}(t))  \triangleq &~ \mathbb{E} \big\{ L(\boldsymbol{Q}(t+1)) - L(\boldsymbol{Q}(t)) \big\vert \boldsymbol{Q}(t)\big\}\\
		& + V \mathbb{E} \left\{ m(t) + \gamma g(t) \big \vert \boldsymbol{Q}(t) \right\},
	\end{split}
\end{equation}
where $V$ is a positive constant that determines the balance between queueing stability and minimizing total system cost. 

To transform problem (\ref{problem_long_term}) into (\ref{problem p2}), we apply \ref{def-Lfunc} and \ref{drift-plus-penalty} and we have
%
	\begin{equation}\label{drift_plus_penalty_expand1}
	\arraycolsep=1.1pt\def\arraystretch{1.5}
		\begin{array}{cl}
			& \displaystyle \Delta_V (\boldsymbol{Q}(t)) \\
			= & \displaystyle \frac{1}{2} 
			\mathbb{E}\Big\{ \sum_{k=1}^{K}\left[Q_k^p(t+1)\right]^2 - \sum_{k=1}^{K}
				\left[Q_k^p(t)\right]^2 \Big\vert \boldsymbol{Q}(t) \Big\} \\
			+ & \displaystyle 
			\frac{\alpha}{2} \sum_{s\in\mathcal{S}}
			\mathbb{E}\Big\{ 
			\sum_{f \in \mathcal{F}_{s} \bigcap \mathcal{F}_{in}}
			\Big[
			\left(Q_f^s(t+1)\right)^2 - \left(Q_f^s(t)\right)^2 
			\Big]
			\Big\vert \boldsymbol{Q}(t) \Big\} \\
			+ & \displaystyle 
			\frac{\alpha}{2} \sum_{s\in\mathcal{S}}
			\mathbb{E}\Big\{ 
			\sum_{f \in \mathcal{F}_{s} \backslash \mathcal{F}_{in}}
			\Big[
			\left(Q_f^s(t+1)\right)^2 - \left(Q_f^s(t)\right)^2 
			\Big]
			\Big\vert \boldsymbol{Q}(t) \Big\} \\		
			+ & \mathbb{E} \left\{ Vm(t) + V\gamma g(t) \bigg \vert \boldsymbol{Q}(t) \right\}.
		\end{array}
	\end{equation}
By (\ref{q_update_pred}) - (\ref{q_update_non_ingr}), it follows that
	
	1) for $k \in \{1,\dots,K\}$,
	\begin{equation} \label{q_update_bound_pred}
	\arraycolsep=1.1pt\def\arraystretch{1.5}
	\begin{array}{lcl}
	 	\displaystyle 
	 	\left[ Q^p_k(t+1) \right]^2 & 
	 	\leq & \displaystyle \left[ Q^{p}_{k}(t) \right]^2 + \left[ A_k(t+D_k+1) \right]^2 
		 	+ \left[ \mu_k(t) \right]^2
	 	\\
	 	& & \displaystyle + 2Q^p_k(t)\left[ A_k(t+D_k+1)-\mu_k(t)\right],
	 \end{array}
	\end{equation}
	where we define $\mu_k(t) \triangleq \sum_{s \in \mathcal{S}_{f}} \mu^{s}_{k}(t)$.
	
	2) for $s \in \mathcal{S}$ and $f \in \mathcal{F}_{s} \bigcap \mathcal{F}_{in}$,
	\begin{equation}\label{q_update_bound_ingr}
		\arraycolsep=1.1pt\def\arraystretch{1.3}
		\begin{array}{rl}
			   \displaystyle \left[ Q^s_f(t+1)\right]^2
			   \leq & \displaystyle 
			   \left[ Q^{s}_{f}(t)\right]^2 + 
			   \left[\mu^{s}_{k_{f}}(t) \right]^2 +
			   \left[ \phi_f(Y^{s}_{f}(t)) \right]^2 \\
			   & + 2Q^s_f(t)
			   \left[ \mu^{s}_{k_{f}}(t) - \phi_{f}(Y^{s}_{f}(t))\right].
		\end{array}
	\end{equation}
	
	3) for $s \in \mathcal{S}$ and $f \in \mathcal{F}_{s} \backslash \mathcal{F}_{in}$,
	\begin{equation}\label{q_update_bound_non_ingr}
	\arraycolsep=1.1pt\def\arraystretch{1.5}
	\begin{array}{rl}
		   & \displaystyle 
		   \left[ Q^s_f(t+1)\right]^2
			 \\
		\leq & \displaystyle 
			   \left[ Q^{s}_{f}(t)\right]^2 + 
			   \Big[
			   	\sum_{s' \in \mathcal{S}_{p(f)}} 
			   	X^{(s', s)}_{p(f)}(t) B^{s'}_{p(f)}(t) 
			   \Big]^2 +
			   \Big[ \phi_f(Y^{s}_{f}(t)) \Big]^2 \\	   
		& \displaystyle 
			   + 2Q^s_f(t) 
			   \Big[ 
			   \sum_{s' \in \mathcal{S}_{p(f)}} 
			   	X^{(s', s)}_{p(f)}(t) B^{s'}_{p(f)}(t)
			   - \phi_{f}(Y^{s}_{f}(t))
			   \Big].
	\end{array}
	\end{equation}
	
	By (\ref{drift-plus-penalty}) -- (\ref{q_update_bound_non_ingr}) and the boundedness of request arrival numbers, service capacities on switches and controllers, and according to (\ref{constraint_mu}), 
we have $\mu^{s}_{k}(t) \leq
	        \le (D+1) \cdot a_{max}$ for $k \in \{1, \dots, K\}$, and
	        
\begin{equation}\label{drift_plus_penalty_expand3}
	\arraycolsep=1pt\def\arraystretch{1.5}
	    \begin{array}{rl}
			\displaystyle \Delta_V (\boldsymbol{Q}(t)) \leq & \displaystyle
	        B +
				\mathbb{E}\bigg\{
					\sum_{k=1}^{K} Q_{k}^p(t) 
					\left[ A_k(t+D_k+1) -  \mu_k(t) \right]  
					\bigg\vert \boldsymbol{Q}(t) \bigg\} \\ 
			& \displaystyle  \!\!\!\!\!\!\!\!
			+ \alpha
				\mathbb{E}\bigg\{
					\sum_{s \in \mathcal{S}} \sum_{f \in \mathcal{F}_{s} \bigcap \mathcal{F}_{in}} 
					\!\!\!\!\!\!\!Q^s_f(t)
			   \left[ \mu^{s}_{k_{f}}(t) - \phi_{f}(Y^{s}_{f}(t))\right]
					\bigg\vert \boldsymbol{Q}(t) \bigg\}\\
		 	& \displaystyle \!\!\!\!\!\!\!\!
		 	+ \alpha
				\mathbb{E}\bigg\{
					\sum_{s \in \mathcal{S}} \sum_{f \in \mathcal{F}_{s} \backslash \mathcal{F}_{in}} 
					\!\!\!\!\!Q^s_f(t)
			   \bigg[  
			   \sum_{s' \in \mathcal{S}_{p(f)}} 
			   	X^{(s', s)}_{p(f)}(t) B^{s'}_{p(f)}(t)
			   - \\
			  & \displaystyle 
			  \left. \left.
			  \phi_{f}(Y^{s}_{f}(t))\right]
					\bigg\vert \boldsymbol{Q}(t) \right\} + \displaystyle \mathbb{E} \left\{ Vm(t) + V\gamma g(t) \bigg \vert \boldsymbol{Q}(t) \right\},
	    \end{array}
\end{equation}
where we define $b_{max} \triangleq \max_{f \in \mathcal{F}} |\mathcal{S}_{f}| $ as the maximum number of instances among all VNFs, and the constant $B$ as
	\begin{equation}\label{def_B}
	\arraycolsep=1pt\def\arraystretch{1.8}
	\begin{array}{cl}
		B \triangleq & \displaystyle
	        \frac{1}{2} 
	        \left[ 
	        	K (a_{max})^{2} + K (D+1)^2 (a_{max})^2 
	        \right] \\
	        & \displaystyle
	        + \frac{\alpha}{2} K b_{max}
	        \left[
	        	 (D+1)^2 (a_{max})^2 + (\phi_{max})^2
	        \right] \\
			& \displaystyle
	        + \frac{\alpha}{2} K b_{max}
	        \left[
	        	(b_{max})^2 (\phi_{max})^2 + (\phi_{max})^2
	        \right]. \\
	\end{array}
\end{equation} 
By substituting (\ref{def_B}) into (\ref{drift_plus_penalty_expand3}),
canceling the terms irrelevant to the decision variables $\boldsymbol{\mu}(t)$, $\boldsymbol{X}(t)$, and $\boldsymbol{Y}(t)$, 
we obtain
\begin{equation}\label{drift_plus_penalty_expand5}
	    \begin{split} 
	        \Delta_V (\boldsymbol{Q}(t)) & \leq B + C(\boldsymbol{Q}(t)) - 
	        \mathbb{E}\left\{
					\sum_{k=1}^{K} Q_{k}^p(t)
					\mu_k(t)
					\bigg\vert \boldsymbol{Q}(t) \right\}
					\\
			+ & \displaystyle \alpha
				\mathbb{E}\bigg\{
					\sum_{s \in \mathcal{S}} \sum_{f \in \mathcal{F}_{s} \bigcap \mathcal{F}_{in}} 
					\!\!\!\!\!\!\!Q^s_f(t)
			   \left[ \mu^{s}_{k_{f}}(t) - \phi_{f}(Y^{s}_{f}(t))\right]
					\bigg\vert \boldsymbol{Q}(t) \bigg\}\\
			+ & \displaystyle \alpha
				\mathbb{E}\bigg\{
					\sum_{s \in \mathcal{S}} \sum_{f \in \mathcal{F}_{s} \backslash \mathcal{F}_{in}} 
					\!\!\!\!\!Q^s_f(t)
			   \bigg[ 
			   \sum_{s' \in \mathcal{S}_{p(f)}} 
			   	X^{(s', s)}_{p(f)}(t) B^{s'}_{p(f)}(t)
			  \\
			  & \displaystyle
			  \left. \left.
			    - \phi_{f}(Y^{s}_{f}(t))\right]
					\bigg\vert \boldsymbol{Q}(t) \right\} +
			V \mathbb{E} \left\{ m(t) + \gamma g(t) \bigg \vert \boldsymbol{Q}(t) \right\},
		\end{split}
	\end{equation}
	where $C(\mathbf{Q}(t))$ includes the terms only related to $\mathbf{Q}(t)$.
Next, recalling (\ref{def_comm_cost_total}) and (\ref{def_comp_cost_total}), we have
\begin{equation}\label{drift_plus_penalty_expand8}
	    \begin{split}
	        \displaystyle &\Delta_V (\boldsymbol{Q}(t)) \leq B + C(\boldsymbol{Q}(t)) - 
	        \mathbb{E}\left\{
					\sum_{k=1}^{K} Q_{k}^p(t) \mu_k(t)
					\bigg\vert \boldsymbol{Q}(t) \right\}
					\\
			+ & \displaystyle \alpha
				\mathbb{E}\bigg\{
					\sum_{s \in \mathcal{S}} \sum_{f \in \mathcal{F}_{s} \bigcap \mathcal{F}_{in}} 
					\!\!\!\!\!\!\!Q^s_f(t)
			   \left[ \mu^{s}_{k_{f}}(t) - \phi_{f}(Y^{s}_{f}(t))\right]
					\bigg\vert \boldsymbol{Q}(t) \bigg\}\\
			+ & \displaystyle \alpha
				\mathbb{E}\bigg\{
					\sum_{s \in \mathcal{S}} \sum_{f \in \mathcal{F}_{s} \backslash \mathcal{F}_{in}} 
					\!\!\!\!\!Q^s_f(t)
			   \bigg[ 
			   \sum_{s' \in \mathcal{S}_{p(f)}} 
			   	X^{(s', s)}_{p(f)}(t) B^{s'}_{p(f)}(t)
			  \\
			- & \displaystyle
			    \phi_{f}(Y^{s}_{f}(t))\bigg]
					\bigg\vert \boldsymbol{Q}(t) \bigg\} 
			+ 	        V\gamma \mathbb{E}\bigg\{ 
	        		\sum_{s \in \mathcal{S}} \sum_{f \in \mathcal{F}_{s}}
		\boldsymbol{\lambda}_s^{T} Y^{s}_{f}(t)
	        \bigg\vert \boldsymbol{Q}(t) \bigg\} 
	        \\
	    	+ &
			\displaystyle
	        V \mathbb{E}\bigg\{ 
	        	\sum_{(s', s) \in \mathcal{E}} \!
	        	\sum_{f \in \mathcal{F}_{nt}} 
    		B^{s'}_{f}(t) 
    		X^{(s',s)}_{f}(t) w_{s', s}(t)  
		        \bigg\vert \boldsymbol{Q}(t) \bigg\}. \\
	    \end{split}
\end{equation}
Note that we have
\begin{equation}
	\begin{array}{rl}
		& \displaystyle
		\sum_{s \in \mathcal{S}} \sum_{f \in \mathcal{F}_{s} \bigcap \mathcal{F}_{in}} 
					\!\!\!\!\!\!\!Q^s_f(t) \mu^{s}_{k_{f}}(t) 
		=
		\sum_{k=1}^{K} \sum_{s \in \mathcal{S}_{f_{k,1}}} 
		Q_{f_{k,1}}^{s}(t) \mu^{s}_{k}(t).
		\\
	\end{array}
\end{equation}
and
\begin{equation}
	\begin{array}{cl}
		& \displaystyle 
		\sum_{s \in \mathcal{S}} 
		\sum_{f \in \mathcal{F}_{s} \backslash \mathcal{F}_{in}} 
					\!\!\!\!\!
					Q^s_f(t) 
					\sum_{s' \in \mathcal{S}_{p(f)}} 
					X^{(s', s)}_{p(f)}(t) B^{s'}_{p(f)}(t)
		\\
		= & \displaystyle 
		\sum_{(s', s) \in \mathcal{E}} 
		\sum_{f \in \mathcal{F}_{nt}}
		Q^{s}_{n(f)}(t) 
		X^{(s', s)}_{f}(t) B^{s'}_{f}(t), \\
	\end{array}
\end{equation}
where indicator $\mathcal{I}_{p(f) \in \mathcal{F}_{s'}}\!=\!1$ if $p(f) \in \mathcal{F}_{s'}$ and zero otherwise. Note that the last two equality holds because if $p(f) \notin \mathcal{F}_{s'}$, then $X^{(s',s)}_{f}(t) = 0$ for all the time.
Hence, we have
\begin{equation}\label{drift_plus_penalty_expand8}
	    \begin{split}
	        \displaystyle &\Delta_V (\boldsymbol{Q}(t)) \leq B + C(\boldsymbol{Q}(t))
			+ \mathbb{E}\Big\{J_t(\mathbf{\mu}, \mathbf{X}, \mathbf{Y}) \Big\vert \boldsymbol{Q}(t) \Big\},
		\end{split}
\end{equation}
in which we define
\begin{equation}\label{def: j_t}
	    \begin{split}
			& J_t(\mathbf{\mu}, \mathbf{X}, \mathbf{Y}) \\ 
			\triangleq
			& \displaystyle
				\mathbb{E}\Big\{
					\sum_{k=1}^{K}  
					\sum_{s \in \mathcal{S}_{f_{k,1}}}
					\Big[  
						- Q^{p}_{k}(t) + 
						\alpha 
						Q_{f_{k,1}}^{s}(t)
					\Big]  \mu^{s}_{k}(t)
					\Big\vert \boldsymbol{Q}(t) \Big\} \nonumber
		\end{split}
\end{equation}
\begin{equation}
	\begin{split}		
			+ & \displaystyle
				\mathbb{E}\bigg\{\!\!\!\!
					\sum_{(s', s) \in \mathcal{E}}  
					\sum_{f \in \mathcal{F}_{nt}} 
					\!\!\!\!\!
					\left[ \alpha Q^{s}_{n(f)}(t) + V w_{s', s}(t) 
					   \right] \cdot
			   		X^{(s', s)}_{f}(t) B^{s'}_{f}(t)
					\bigg\vert \boldsymbol{Q}(t) \bigg\}\\
	        + & \displaystyle
	        \mathbb{E}\bigg\{ 
	        		\sum_{s \in \mathcal{S}} \sum_{f \in \mathcal{F}_{s}}
	        		\!\!\!\!
	        		\left[
						V \gamma \boldsymbol{\lambda}^{T}_s Y_{f}^{s}(t) 
						- \alpha Q^{s}_{f}(t) \phi_f(Y^{s}_{f}(t))
					\right]
	        \bigg\vert \boldsymbol{Q}(t) \bigg\}. \\
	    \end{split}
\end{equation}
By (\ref{drift_plus_penalty_expand8}), to optimize the long-term time-average of total system cost while stabilizing all queues, we switch to minimizing the upper bound of the drift-plus-penalty term, so as to direct the drift and objective function value in a negative direction towards zero. In this way, we can transform problem \textbf{P1} into a series of subproblems over time slots and solve them on a per-time-slot basis.
Specifically, we first denote the optimal solutions to minimize $J_t$ during time slot $t$ by $\boldsymbol{\mu^*}$, $\boldsymbol{X^*}$, and $\boldsymbol{Y^*}$. 
Accordingly, for any other feasible scheduling decisions $\boldsymbol{\mu}$, $\boldsymbol{X}$, and $\boldsymbol{Y}$ made during time slot $t$, we have that
\begin{equation}
	J_t(\boldsymbol{\mu^*}, \boldsymbol{X^*}, \boldsymbol{Y^*}) \leq 
	J_t(\boldsymbol{\mu}, \boldsymbol{X}, \boldsymbol{Y})),\,w.p.1.
\end{equation}
By taking the conditional expectation on both sides on $\boldsymbol{Q}(t)$, we have
\begin{equation}\label{cond-expec ineq}
	\mathbb{E}\Big\{
	J_t(\boldsymbol{\mu^*}, \boldsymbol{X^*}, \boldsymbol{Y^*})\Big\vert \boldsymbol{Q}(t) \Big\}
	\leq 
	\mathbb{E}\Big\{ 
	J_t(\boldsymbol{\mu}, \boldsymbol{X}, \boldsymbol{Y}))\Big\vert \boldsymbol{Q}(t) \Big\}.
\end{equation}
By (\ref{cond-expec ineq}), 
we know that $\boldsymbol{\mu^*}$, $\boldsymbol{X^*}$, and $\boldsymbol{Y^*}$ also minimize the conditional expectation of $J_t(\boldsymbol{\mu}, \boldsymbol{X}, \boldsymbol{Y}))$, 
thus minimizing the upper bound of the drift-plus-penalty term in (\ref{drift_plus_penalty_expand8}). 
This implies that, instead of directly solving the long-term stochastic optimization problem \textbf{P1}, we can opportunistically solve the following problem during each time slot $t$.
	\begin{equation}
	\begin{split}
	    & J_t(\boldsymbol{\mu}, \boldsymbol{X}, \boldsymbol{Y}) \triangleq \sum_{k=1}^{K}  
					\sum_{s \in \mathcal{S}_{f_{k,1}}}
					\left[  
						- Q^{p}_{k}(t) + 
						\alpha 
						Q_{f_{k,1}}^{s}(t)
					\right]  
					\mu^{s}_{k} \\
		& + 
	    \sum_{(s', s) \in \mathcal{E}}  
					\sum_{f \in \mathcal{F}_{nt}}
	    \left[
	    Vw_{s',s}\left(t\right) 
	    + \alpha Q_{f}^{s}\left(t\right)
	    \right]
	    		B^{s'}_{p(f)}(t)X^{(s', s)}_{f} 
	    \\
	    & +
	    \sum_{s\in\mathcal{S}}
	    \sum_{f\in\mathcal{F}_{s}}
	    \left[
	    V\boldsymbol{\gamma}_{s}^{T}Y_{f}^{s}
	    -
	    \alpha Q_{n(f)}^{s}\left(t\right)\phi_{f}\left(Y_{f}^{s}\right) 
	    \right]. 
	    \end{split}
	    \end{equation}
	    \IEEEQED

	\section*{Appendix-B\\
	Proof of Theorem 1}
	We assume there is an S-only algorithm \cite{neely2010stochastic} which achieves the optimal time-average total cost (infimum) $m^* + \gamma g^*$ with action $\tilde{\boldsymbol{\mu}}(t)$, $\tilde{\boldsymbol{X}}(t)$, and $\tilde{\boldsymbol{Y}}(t)$, 
	for $t= \{0,1,2,\cdots \}$. 
	From \cite{neely2010stochastic}, we know that to ensure queue stability, the expectation of arrival rate should be no more than the expectation of service rate, \textit{i.e.}, there exists some $\epsilon \geq 0 $ such that,
	\begin{equation}\label{S-only inequality}
	    \begin{array}{cl}
	        & \displaystyle 
	        \mathbb{E} \bigg\{ \sum_{k=1}^{K} A_{k}(t) \big \vert \mathbf{Q}(t) \bigg\} + \epsilon 
	        \leq  
	        	\mathbb{E} \bigg\{ 
	        	\sum_{f \in \mathcal{F}} 
	        	\sum_{s \in \mathcal{S}_{f}} \phi_{f}(\tilde{Y}^{s}_{f}(t))
	        	\big \vert \mathbf{Q}(t)\bigg\}.
	    \end{array}
	\end{equation}
	We denote $\boldsymbol{\mu}'(t)$ $\boldsymbol{X}'(t)$, $\boldsymbol{Y}'(t)$ as the decisions within time $t$, and $m'(t)$, $g'(t)$ as the cost incurred by P-OSCARS. By (\ref{drift_plus_penalty_expand5}), the one-slot drift-plus-penalty $\Delta_V(\mathbf{Q}(t))$ is given by
	\begin{equation}
	    \begin{split}
	        & \mathbb{E} \left\{ L(\mathbf{Q}(t+1)) - L(\mathbf{Q}(t)) \big\vert \mathbf{Q}(t)\right\} + V \mathbb{E} \left\{ m'(t) + \gamma g'(t) \big\vert \mathbf{Q}(t)\right\}
	        \\
	        \leq & B + C(\mathbf{Q}(t)) - 
	        \mathbb{E}\bigg\{
					\sum_{k=1}^{K} Q_{k}^p(t) \mu^{\prime}_k(t)
					\Big\vert \boldsymbol{Q}(t) \bigg\}\\
	    	+ & \displaystyle \alpha
				\mathbb{E}\bigg\{
					\sum_{s \in \mathcal{S}} \sum_{f \in \mathcal{F}_{s} \bigcap \mathcal{F}_{in}} 
					\!\!\!\!\!\!\!Q^s_f(t)
			   \left[ {\mu^{\prime}}^{s}_{k_{f}}(t) - \phi_{f}({Y^{\prime}}^{s}_{f}(t))\right]
					\bigg\vert \boldsymbol{Q}(t) \bigg\}\\
			+ & \displaystyle \alpha
				\mathbb{E}\Big\{
					\sum_{s \in \mathcal{S}} \sum_{f \in \mathcal{F}_{s} \backslash \mathcal{F}_{in}} 
					\!\!\!\!\!Q^s_f(t)
			   \Big[ 
			   		\sum_{s' \in \mathcal{S}_{p(f)}} 
			   		{X^{\prime}}^{(s', s)}_{f}(t) 
			   		B^{s'}_{p(f)}(t) - 
			  \\
			  & \displaystyle
			  \phi_{f}({Y^{\prime}}^{s}_{f}(t))\Big]
					\big\vert \boldsymbol{Q}(t) \Big\} + 
			V \mathbb{E} \Big\{ m'(t) + \gamma g'(t) \Big \vert \boldsymbol{Q}(t) \Big\}.  
	    \end{split}
	\end{equation}
	Recall that $m^*$ and $g^*$ are the infimum of the achievable time-average total cost.
	By substituting (\ref{S-only inequality}) into the above expression and adopting the definition (\ref{def_total_q}), we obtain
	\begin{equation}\label{proof_step2}
	    \begin{split}
	        & \mathbb{E} \left\{ L(\mathbf{Q}(t+1)) - L(\mathbf{Q}(t)) \big\vert \mathbf{Q}(t)\right\} + V \mathbb{E} \left\{ m'(t) + \gamma g'(t) \big\vert \mathbf{Q}(t)\right\}\\
	        & \leq B + V \mathbb{E} \left\{ m^*(t) + \gamma g^*(t) \big\vert \mathbf{Q}(t)\right\}- \epsilon \mathbb{E} \left\{ h(t) \big\vert \mathbf{Q}(t)\right\}.
	    \end{split}
	\end{equation}
By taking expectation over both sides and
	summing over $t \in \{0, 1, \cdots, T-1\}$, we have 
	\begin{equation}\label{proof_step4}
	    \begin{array}{l}
	    	\displaystyle
	        \mathbb{E} \left\{ L(\mathbf{Q}(T-1)) \right\} - \mathbb{E} \Big\{ L(\mathbf{Q}(0))\Big\} + V \mathbb{E} \left\{ \sum_{t=0}^{T-1}\left( m'(t) + \right.\right. \\
	        \displaystyle
	        \left.\left.
	        \gamma g'(t) \right)\right\} 
	        \leq BT\! +\! V \mathbb{E} \Big\{\sum_{t=0}^{T-1}\left( m^*(t) + \gamma g^*(t) \right) \Big\} \!\! - \! \epsilon \mathbb{E} \left\{ h(t) \right\}.\\
	    \end{array}
	\end{equation}
	By (\ref{proof_step4}), we are ready to show the 
	$[O(V), O(1/V)]$ trade-off.
	
	1) By dividing both sides of (\ref{proof_step4}) by $VT$, rearranging items and neglecting non-positive quantities in right side, we obtain
	\begin{equation}\label{performance_cost1}
	    \begin{split}
	    &\frac{1}{T}\sum_{t=0}^{T-1} \mathbb{E} \left\{ \left( m'(t) + \gamma g'(t) \right) \right\} \\
	    & \leq \frac{B}{V} + \frac{\mathbb{E}\{L(\mathbf{Q}(0))\}}{VT} + \frac{1}{T}\sum_{t=0}^{T-1} \mathbb{E} \left\{ \left( m^*(t) + \gamma g^*(t) \right) \right\}.
	    \end{split}
	\end{equation}
	As $T\rightarrow \infty$, we have
	\begin{equation}\label{performance_cost2}
	    \begin{split}
	    \limsup_{T\rightarrow \infty} \frac{1}{T}&\sum_{t=0}^{T-1} \mathbb{E} \left\{ \left( m'(t) + \gamma g'(t) \right) \right\} \\
	    &\leq \limsup_{T\rightarrow \infty} \frac{1}{T}\sum_{t=0}^{T-1} \mathbb{E} \left\{ \left( m^*(t) + \gamma g^*(t) \right) \right\} + \frac{B}{V}. \\ 
	    \end{split}
	\end{equation}
	
	2) Similarly, by dividing both sides of (\ref{proof_step4}) by $\epsilon T$, we get
	\begin{equation}\label{performance_queue1}
	    \begin{split}
	    &\frac{1}{T}\sum_{t=0}^{T-1} \mathbb{E} \left\{ h(t) \right\} \\
	    &\leq \frac{B}{\epsilon} + \frac{\mathbb{E}\{L(\mathbf{Q}(0))\}}{\epsilon T} + \frac{V\sum_{t=0}^{T-1}\mathbb{E} \left\{ \left( f^*(t) + \gamma g^*(t) \right)\right\}}{\epsilon T}.
	    \end{split}
	\end{equation}
	As $T\rightarrow \infty$, we obtain
	\begin{equation}\label{performance_queue3}
	    \bar{h} \leq \frac{V}{\epsilon} \left({m^*} + \gamma {g^*} \right) + \frac{B}{\epsilon}.
	\end{equation}
	\IEEEQED

\end{document}